\documentclass[useAMS,usenatbib]{mn2e}
\usepackage{natbib}
\usepackage{graphicx}
\usepackage{times}
\usepackage{rotate} 
\usepackage{color,url}
\newcommand{\simgt}{\lower.5ex\hbox{$\; \buildrel > \over \sim \;$}}
\newcommand{\simlt}{\lower.5ex\hbox{$\; \buildrel < \over \sim \;$}}
\newcommand{\ave}[1]{\left\langle #1\right\rangle}

\def\hkpc{h^{-1}{\rm kpc}}
\def\hMpc{h^{-1}{\rm Mpc}}
\def\hGpc{h^{-1}{\rm Gpc}}
\def\hMsun{h^{-1}M_\odot}
\def\Mpch{h~{\rm Mpc}^{-1}}

\begin{document}
\title[Understanding the nature of LRGs]
{Understanding the nature of luminous red galaxies (LRGs): Connecting
LRGs to central and satellite subhalos}

\author[S. Masaki et al.]
{Shogo Masaki$^{1,2}$\thanks{E-mail: shogo.masaki@nagoya-u.jp}\thanks{JSPS Fellow},
Chiaki Hikage$^3$, 
Masahiro Takada$^4$,
David N. Spergel$^{4,5}$,
\newauthor{and 
Naoshi Sugiyama$^{1,3,4}$}\\
$^1$ Department of Physics, Graduate School of Science, Nagoya University,  Aichi 464-8602, Japan\\
$^2$ NTT Secure Platform Laboratories, Tokyo 180-8585, Japan\\
$^3$ Kobayashi Maskawa Institute (KMI), Nagoya University, Aichi 464-8602, Japan\\
$^4$ Kavli Institute for the Physics and Mathematics of the Universe 
(Kavli IPMU, WPI)
, The University of Tokyo, Chiba 277-8582, Japan\\
$^5$ Department of Astrophysical Sciences, Princeton University, Peyton
Hall, Princeton NJ 08544, USA \\
}

\maketitle
\label{firstpage}

%%%%%%%%%%%%%%%%%%%%%%%%%%%%%%%%%%%%%%%%%%%%%%%%%%%%%%%%
\begin{abstract}
We develop a novel abundance matching method to construct a mock catalog of
luminous red galaxies (LRGs) in the Sloan Digital Sky Survey (SDSS), 
using catalogs of halos and subhalos in $N$-body simulations for a
$\Lambda$-dominated, cold dark matter model. Motivated by observations suggesting that
LRGs are passively-evolving, massive early-type galaxies with a typical
age $\simgt 5~$Gyr, we assume that simulated halos at $z=2$ ($z2$-halo)
are progenitors for LRG-host subhalos observed today, 
and  we label the most tightly bound
particles in 
each progenitor $z2$-halo as LRG ``stars''.   
We then identify the subhalos containing
these stars to  $z=0.3$ (SDSS redshift) in
descending order of the masses of $z2$-halos until the comoving number
density of the matched subhalos becomes comparable to the measured
number density of SDSS LRGs, $\bar{n}_{\rm LRG}=10^{-4}~h^3~{\rm
Mpc}^{-3}$. Once the above prescription is determined,  
our only free parameter is the number density of halos
identified at $z=2$ and this parameter is fixed to match
the observed number density at $z = 0.3$.
By tracing subsequent merging and assembly histories of
each progenitor $z2$-halo, we can directly compute, from 
the mock catalog,
the distributions of central and satellite LRGs and their
internal motions
in each host halo at $z=0.3$.  While the SDSS LRGs are galaxies selected
by the magnitude and color cuts from the SDSS images and are not
necessarily a stellar-mass-selected sample, our mock catalog  reproduces a host of 
SDSS measurements: the halo occupation distribution for central and
satellite LRGs, the projected auto-correlation function of LRGs, the
cross-correlation of LRGs with shapes of background galaxies (LRG-galaxy
weak lensing), and the nonlinear redshift-space distortion effect, the
Finger-of-God effect, in the angle-averaged redshift-space power
spectrum. 
The mock catalog generated based on our method can be used for
removing or calibrating systematic errors in the cosmological
interpretation of LRG clustering measurements as well as for
 understanding the nature of LRGs such as their formation and assembly
 histories. 
\end{abstract}
%%%%%%%%%%%%%%%%%%%%%%%%%%%%%%%%%%%%%%%%%%%%%%%%%%%%%%%%


\section{Introduction}

Galaxy redshift surveys are one
of the primary tools for studying the large-scale structure in the
Universe
\citep{DavisHuchra:82,deLapparentetal:86,Kirshneretal:87,SDSS,Peacocketal:01}.
Over the coming decade, astronomers will have even larger surveys
including BOSS\footnote{\url{http://cosmology.lbl.gov/BOSS/}}
\citep{2012arXiv1208.0022D},
WiggleZ\footnote{\url{http://wigglez.swin.edu.au/site/}}
\citep{Blakeetal:11}, VIPERS\footnote{\url{http://vipers.inaf.it/}},
FMOS\footnote{\url{http://www.naoj.org/Observing/Instruments/FMOS/}},
HETDEX\footnote{\url{http://hetdex.org/}},
BigBOSS\footnote{\url{http://bigboss.lbl.gov/}} \citep{BigBOSS}, Subaru
Prime Focus Spectrograph
(PFS\footnote{\url{http://sumire.ipmu.jp/pfs/intro.html}};
\citealt{Ellisetal:12}),
Euclid\footnote{\url{http://sci.esa.int/euclid}}, and
WFIRST\footnote{\url{http://wfirst.gsfc.nasa.gov/}}.  The upcoming
generation of galaxy redshift surveys 
is aimed at 
understanding cosmic acceleration as well as measuring the composition of the
Universe via measurements of both the geometry and the dynamics of
structure formation
\citep[][]{WangSpergelStrauss:99,Eisensteinetal:99,Tegmarketal:04,Coleetal:05}.

On large scales, galaxies trace the underlying distribution of dark
matter, and
their clustering correlation 
is a standard tool to extract cosmological information from the
measurement. 
Because of their relatively high spatial densities
and their intrinsic bright luminosities,
luminous red galaxies (LRGs) are one of the most useful tracers
\citep{Eisensteinetal:01,Wakeetal:06}.  Measurements of the clustering
properties of LRGs have been used to measure the baryon acoustic
oscillation (BAO) scale
\citep{Eisensteinetal:05,Percivaletal:07BAO,Andersonetal:12} as well as
to constrain cosmological parameters
\citep{Tegmarketal:04,Coleetal:05,Reidetal:10,Saitoetal:11}.

Our lack of a detailed understanding of the relationship between galaxies
and their host halos 
complicates
the analysis
of large-scale clustering data.
 The halo occupation distribution (HOD) approach
or the halo model approach has provided a useful, albeit empirical,
approach to relating galaxies to dark matter 
\citep[see e.g.,][for the pioneer works]{PeacockSmith:00,Seljak:00,Scoccimarroetal:01}.
In these approaches, the distribution of halos is first modeled for a
given cosmological model, e.g. by using $N$-body simulations, and then
galaxies of interest are populated into dark matter halos. 
The previous works have shown that, by adjusting the model parameters,
the HOD based model well reproduces the auto-correlation functions of
LRGs measured from the Sloan Digital Sky
Survey\footnote{\url{http://www.sdss.org/}} (SDSS) 
\citep{Zehavi05,Zhengetal:07,Wakeetal:08,ReidSpergel09,Whiteetal:11}.
It has been shown that LRGs reside in massive
halos with a typical mass of a few times $10^{13}~\hMsun$.
However, the HOD method employs several simplified assumptions. For
instance, the distribution of galaxies is assumed to follow that of dark
matter in their host halo and the model assumes a
simple functional form for the HOD. 

An alternative approach is the so-called 
abundance matching method.
The abundance matching method directly connects 
target galaxies to simulated subhalos assuming a tight and physically-motivated 
relation between their properties, e.g., galaxy luminosity 
and subhalo circular velocity, without employing
 any free fitting 
parameters
\citep[e.g.,][]{Kravtsovetal:04,Conroyetal:06,Trujillo-Gomezetal:11,Reddicketal:12,Masakietal:12a,Nuzaetal:12}.
However, it is not still
clear whether the method can simultaneously reproduce different
clustering measurements such as the auto-correlation function and the
galaxy-galaxy weak lensing \citep[][]{NeisteinKhochfar:12}. Most of the
previous studies
use only the auto-correlation function to test their abundance matching model.

In this paper,
we develop an alternative approach to the abundance matching method for
constructing a mock catalog of LRGs.  
Motivated by observations suggesting that  LRGs are passive, massive early-type galaxies, which
are believed to have formed at $z>1$
\citep[][]{Masjedietal:08,CarsonNichol:10,Tojeiroetal:12}, we assume
that the progenitor halos for LRG-host subhalos are formed at
$z=2$.
We identify 
massive halos at this redshift,  define 
the innermost particles of each progenitor halo as hypothetical 
``LRG-star''
particles, follow the star particles to lower redshifts,
and then identify subhalos  at $z=0.3$ containing these star particles.
We adjust the number of halos identified as LRG progenitors at $z=2$
to match the observed  number density of the SDSS LRGs, 
$\bar{n}_{\rm LRG}\simeq 10^{-4}~h^3~{\rm Mpc}^{-3}$
\citep[also see][for a similar-idea based approach when connecting
galaxies to halos]{Conroyetal:08,Seoetal:08}.
With this method, we can directly trace, from the simulations, how each
progenitor halo at $z=2$ experiences merger(s), is destroyed or survives
at lower redshift as well as which progenitor halos become central or
satellite subhalos (galaxies) in each host halo at $z=0.3$. Thus, our method
allows us to include
assembly/merging histories of the
LRG-progenitor halos. Our method is solely based on a mass-selected
sample of progenitor halos at $z=2$ and does not have {\em any} free 
fitting
parameter because the mass threshold is fixed by matching to the
number density of SDSS LRGs.  We compare statistical quantities computed
from our mock catalog with the SDSS measurements: the HOD, the projected
auto-correlation function of LRGs, the LRG-galaxy weak lensing and the
redshift-space power spectrum of LRGs. Even though our method is rather
simple,  we show that our mock catalog
remarkably well reproduces the different measurements simultaneously.

The structure of this paper is as follows. In Section~\ref{sec:method},
we describe our method to generate a mock catalog of LRGs by using
$N$-body simulations for a $\Lambda$-dominated 
cold dark matter ($\Lambda$CDM)  model as well as
the catalogs of halos and subhalos at $z=2$ and $z=0.3$. In
Section~\ref{sec:results}, we show the model predictions on the relation
between LRGs and dark matter, and compare with the SDSS measurements.
Section~\ref{sec:conclusion} is devoted to discussion and conclusion.

\section{Methods}
\label{sec:method}

\subsection{Cosmological $N$-body simulations}

Throughout this paper 
we use two realizations of cosmological $N$-body simulations generated
using
the publicly-available {\it Gadget-2} code \citep{Springel01a,Springel05}.
For each run, we employ a flat $\Lambda$CDM cosmology with
$\Omega_m=0.272, \Omega_b=0.0441, \Omega_\Lambda=0.728,
H_0=100h=70.2~{\rm km~s^{-1}~Mpc^{-1}}, \sigma_8=0.807$ and $n_s=0.961$
using the same parameters and notation as in the
 the {\it WMAP} 7-yr  analysis \citep{Komatsu11}.
Our 
simulation of larger-size box,
which we hereafter call ``L1000'',
employs $1024^3$ dark matter particles in a box of $1~\hGpc$ on a side.
The L1000 simulation allows
for a higher statistical precision 
in measuring the
correlation functions from the mock catalog.
{To test the effect of numerical resolution on our results, we also use}
a  
 higher resolution simulation that employs
$1024^3$ particles in a box of $300~\hMpc$.
{We call the smaller-box simulation ``L300''.}
The mass resolution for the simulations (mass of an $N$-body particle)
is $7\times10^{10}~\hMsun$ 
or $1.9\times10^9~\hMsun$ for L1000 or L300, respectively.
The initial conditions for both the
simulation runs are generated using the second-order Lagrangian perturbation theory \citep{Crocceetal:06,Nishimichietal:09}
and an  initial matter power 
spectrum  at $z=65$, computed from the {\it CAMB} code \citep{Lewis00}.
We set the gravitational softening parameter to be $30$ and $8~\hkpc$
for the L1000 and L300 runs, respectively. 
 
We use
the friends-of-friends (FoF) group finder \cite[e.g.,][]{Davisetal:85}
with a linking length of $0.2$ in units of the mean interparticle
spacing to create a catalog of halos from the simulation output and use
the {\it SubFind} algorithm  \citep{Springel01b}  to
identify subhalos within each halo. 
 In this paper, we use halos and subhalos that contain more
than 20 particles.
Each particle in a halo region is assigned to either a smooth
component of the parent halo
or to a satellite subhalo, where the smooth
component contains the majority of $N$-body particles in the halo region.
Hereafter we 
call
the smooth component 
a central subhalo and call the 
subhalo(s) 
satellite subhalo(s).
For
each subhalo, we estimate its mass by counting the bounded
particles, 
which we call 
the subhalo mass ($M_{\rm sub}$).
We store the position and velocity data of particles in halos
and subhalos at different redshifts.  To estimate
the virial mass ($M_{\rm vir}$) for each parent halo, we apply the spherical
overdensity method to the FoF halo, where the spherical boundary
region is determined by the interior virial overdensity, $\Delta_{\rm vir}$, 
relative to the mean mass density  \citep{BryanNorman:98}. 
The overdensity $\Delta_{\rm vir}\simeq 268$ at $z=0.3$ for the assumed
cosmological model.  
The virial
radius is estimated from the estimated mass as $R_{\rm vir}=(3M_{\rm
vir}/4\pi\bar{\rho}_{m0}\Delta_{\rm vir})^{1/3}$, where
$\bar{\rho}_{m0} $ is  the comoving matter density.

\subsection{Mock catalog of LRGs: connecting halos at $z=2$
to central and satellite subhalos at $z=0.3$}
\label{sec:mock}

\begin{table*}
\begin{center}
\begin{tabular}{l||llllll}
\hline\hline
Type of LRG-host halos & Total number & $\bar{M}_{\rm vir}~[10^{13}\hMsun]$ 
& $\bar{R}_{\rm vir}~ [\hMpc]$ & $f_{\rm
 sat-LRG}$ & $q_{\rm cen}^{\rm BLRG}$
&$q_{\rm cen}^{\rm FLRG}$
\\ 
\hline
All LRG-host halos & ~ & ~ & ~ & ~ & ~ & ~\\
~~~L1000 & 91,090 & 
$5.64\pm0.11$ 
& $0.804 \pm 0.004$ & $0.0988 \pm 0.0054$ & $0.959 \pm 0.004$ & --\\
~~~L300 & 2,403 
& $5.66$ 
& $0.806$ & $0.119$ & $0.956$ & --\\
%~~~Hikage et al.(2012a) & ~ & ~ & ~ & ~ & ~ & ~\\
Single-LRG halos & ~ & ~ & ~ & ~ & ~ & ~\\
~~~L1000 & 83,891 
& $4.81\pm0.08$ 
& $0.776 \pm 0.004$ & $0.0215 \pm 0.0028$ & $0.979 \pm 0.003$ & -- \\
~~~L300 & 2,166 
& $4.74$ 
& $0.772$ & $0.0226$ & $0.977$ & --\\
~~~Hikage et al.~(2012a) & 87,889 
& $3.7\pm0.4$ 
& $0.77\pm0.03$ & $0.24\pm0.18$ & $0.76\pm0.18$ & --\\
Multiple-LRG halos & ~ & ~ & ~ & ~ & ~ & ~\\
~~~L1000 & 7,199 
& $15.2\pm0.8$ 
& $1.13 \pm 0.02$ & $1.00$ & $0.735 \pm 0.025$ & $0.207 \pm 0.023$ \\
~~~L300 & 237 
& $14.0$ 
& $1.11$ & $1.00$ & $0.764$ & $0.186$ \\
~~~Hikage et al.~(2012a) & 4,157 
& $14.6\pm1.1$ 
& $1.21\pm0.03$ & $1.00$ & $0.63\pm0.21$ & $0.24\pm0.13$\\
\hline\hline
\end{tabular}
\caption{Summary of properties of LRG-host halos, computed from the mock
 LRG catalog in the L1000 and L300 runs (see text for details).
 Here we consider all LRG-host halos and the
 single- and multiple-LRG halos that host only one or multiple LRG(s)
 inside, respectively. 
$\bar{M}_{\rm vir}$ and $\bar{R}_{\rm vir}$ are
 the average virial mass and radius of the host halos (without any
 weight).  $f_{\rm
 sat-LRG}$ is the fraction of halos that have satellite LRG(s) among all the 
 LRG-host halos in either single- or multiple-LRG halos (each row).
 Note that $f_{\rm sat-LRG}$ for multiple-LRG systems 
is
unity by definition
since the halos 
 have satellite LRG(s).
 $q_{\rm cen}^{\rm BLRG}$ is the fraction of
 halos that host its BLRG as a central galaxy among all the host halos,
where BLRG is the brightest LRG, the most massive LRG-progenitor halo at
 $z=2$, compared to other LRG-subhalo(s) in the same host halo at $z=0.3$. 
 Note that, for the single-LRG hosts, we call
 the LRG as the BLRG. $q_{\rm cen}^{\rm FLRG}$ is the fraction of halos that
 host FLRG as a central LRG, where FLRG is the faintest LRG, the
 smallest LRG-progenitor halo, in each multiple-LRG halo. 
 The error bars quoted for
the L1000 mock are the standard deviation 
computed from the 27 divided sub-volumes of L1000 mock each of which has
volume of $333^3~[\hMpc]^3$. Hence, the L1000 results with 
the errors in each row can be compared with the L300 mock results, which has 
comparable volume of
 $300^3~[\hMpc]^3$. For comparison, we also quote the measurement
 results derived from the SDSS DR7 LRG catalog in \citet{Hikage12b},
 where the error bars are $\pm 68\%$ confidence ranges (see text for details).
\label{tab:lrgs}}
\end{center}
\end{table*}

\begin{figure*}
\begin{center}
\includegraphics[width=18cm]{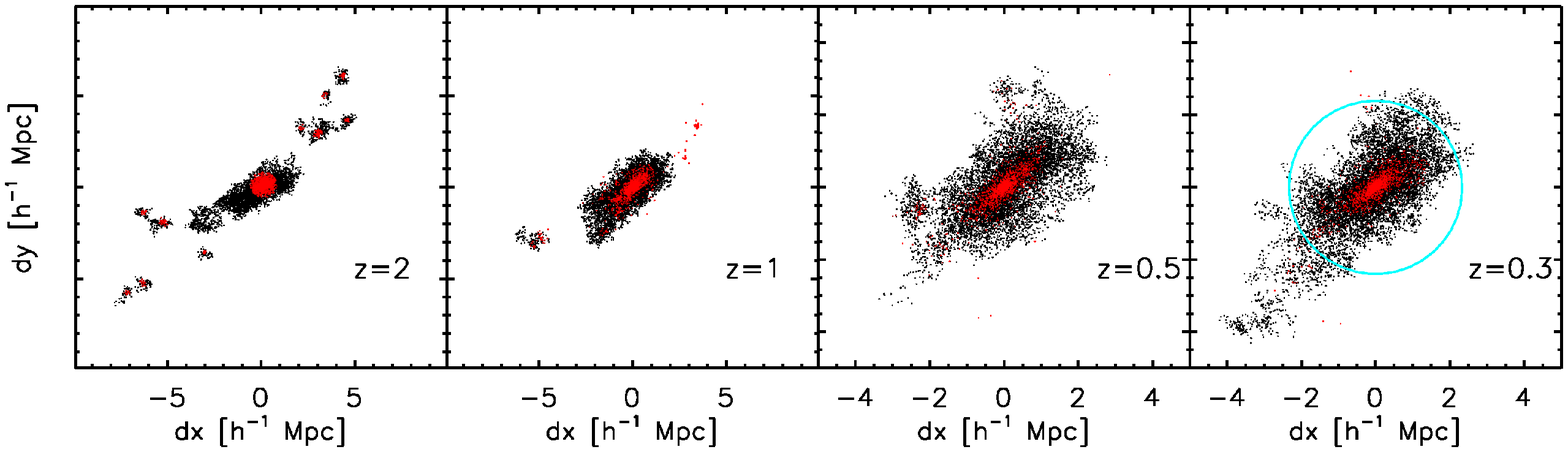}
\includegraphics[width=18cm]{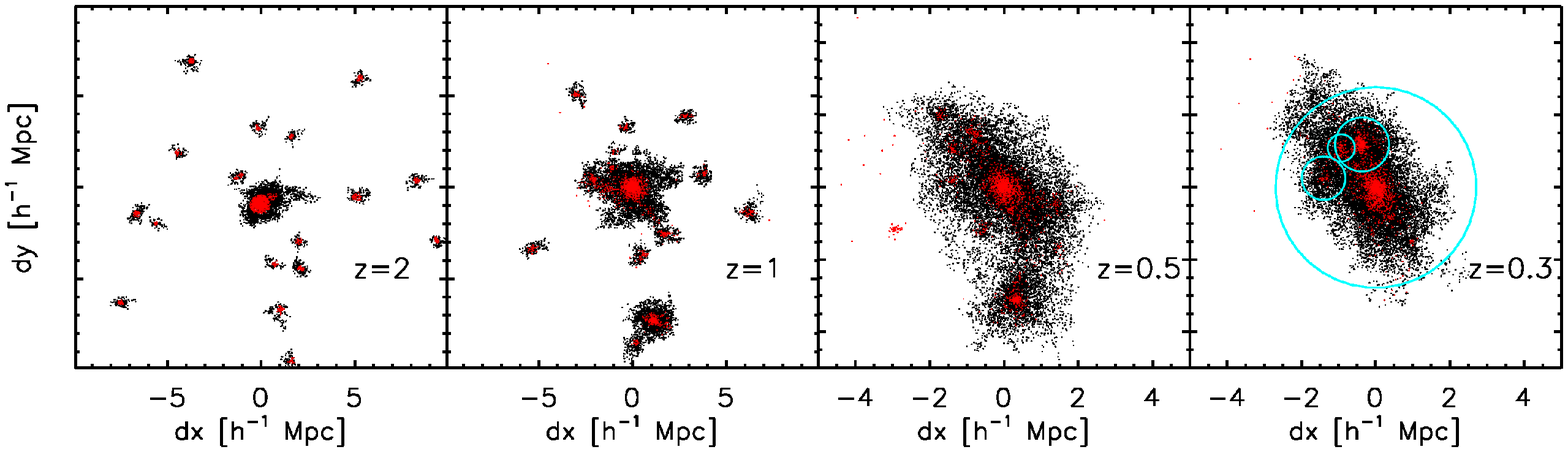}
\caption{Evolution of dark matter ($N$-body) particle distribution
around the region
 of subhalos
 hosting mock LRGs at $z=0.3$, 
taken from our L1000 simulation run. The
upper-row panels are for the region around the host-halo of the brightest LRG 
among single-LRG systems (the host halo mass $M_{\rm vir}=8.42\times 10^{14}~\hMsun$), systems which host only single LRG inside in
the $z=0.3$ output, while the lower-row panels are  the most
massive host-halo among systems hosting one central and three 
satellite LRGs ($M_{\rm vir}=1.44\times10^{15}~\hMsun$). 
The dot symbols in each panel are member particles in the halo regions
at $z=2$ or the subhalo region(s) at lower redshifts.
The red-color particles are 30\% innermost particles of 
each halo at $z=2$ and
selected based on our abundance matching method between the progenitor
halos and the LRG-host subhalos at $z=0.3$ to reproduce the observed
number density of SDSS LRGs (see text for details).  
Then we trace where
the red-color particles are distributed in each subhalo region at lower
redshift.  By matching the red-color particles to central and satellite
subhalos in each host halo of $z=0.3$ output, we can define locations of
each LRG in a host halo at $z=0.3$; if a subhalo at $z=0.3$ contains
more than 50\% of the red-color particles of a progenitor halo, we
define it as an LRG-host subhalo. The upper-row panels show the case
that 11 progenitor halos of LRGs are formed at $z=2$, and then are
merged at lower redshift, forming one central LRG in the host halo at
$z=0.3$. The lower-row panels show that 24 progenitor halos at $z=2$
form one central LRG and three satellite LRGs in the host-halo at
$z=0.3$. 
The blue circles in the panel of $z=0.3$ shows the positions of mock LRGs.
The size of each circle is proportional to $M_{\rm sub}^{1/3}$, 
where $M_{\rm sub}$ is the subhalo mass.
\label{fig:snapshot} }
\end{center}
\end{figure*}
\begin{figure*}
\begin{center}
\includegraphics[width=8.5cm]{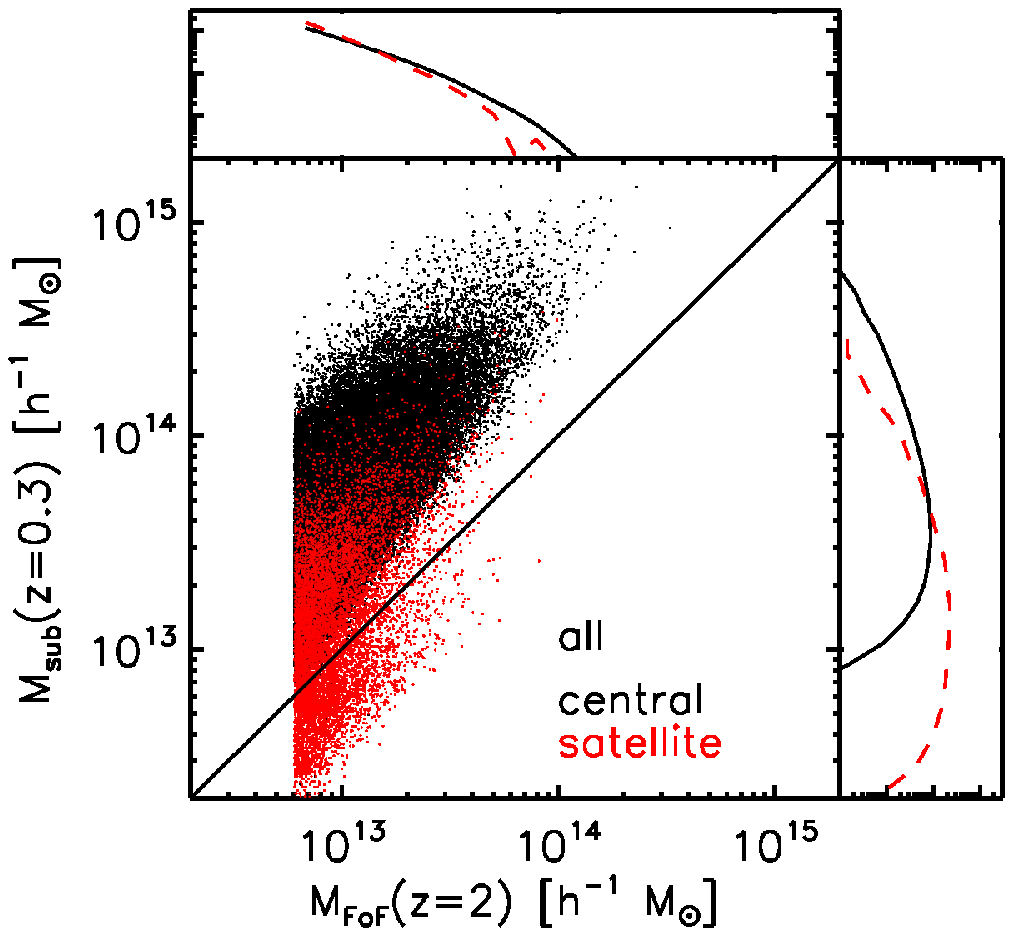}
\includegraphics[width=8.5cm]{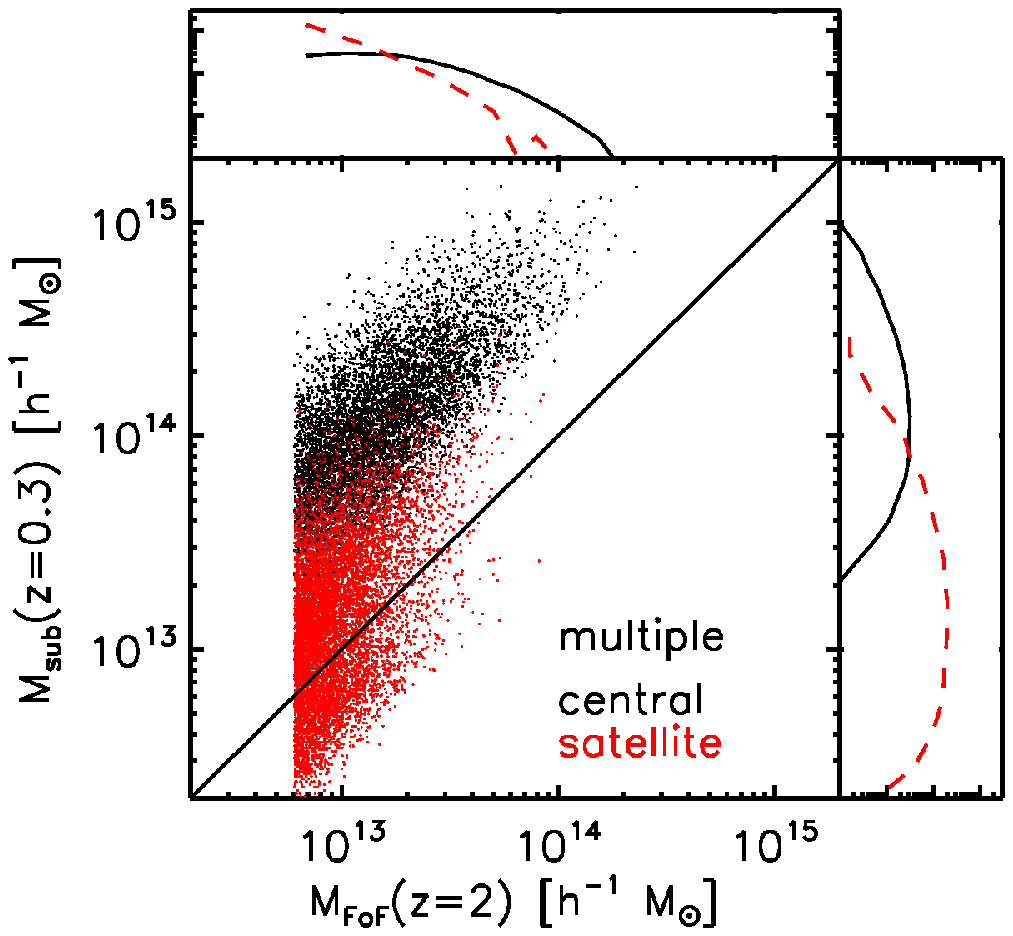}
\caption{Comparison between masses of the LRG-progenitor halos at $z=2$
and the LRG-host subhalos at $z=0.3$, computed from the L1000 run, where
each progenitor halo and subhalo are matched based on our method (see
Figure~\ref{fig:snapshot}). The left and right panels show the results
for all the LRG-host halos and the multiple-LRG systems, respectively.
The black and red points are for central and satellite LRGs,
respectively. Note that the central LRG-subhalo is a smooth component of
the parent halo at $z=0.3$. The line in each panel shows the case that
the progenitor halo does not either gain or lose its mass at $z=0.3$:
$M_{{\rm sub}}(z=0.3)=M_{\rm FoF}(z=2)$.  The figure shows that satellite
LRGs preferentially reside in less massive progenitor halos at $z=2$,
some subhalos for satellite LRGs lose their masses due to tidal
stripping when accreting into more massive halos, and subhalos for
central LRGs gain their masses due to merger. The upper- and right-side
panels in each plot are the projected distributions for central and
satellite LRGs along the $y$- or $x$-axis direction, respectively.
\label{fig:mz2-mz0.3}}
\end{center}
\end{figure*}

LRGs are very useful tracers of large-scale structure as they can
reach a higher redshift, thereby enabling to cover a larger volume with
the spectroscopic survey 
\citep{Eisensteinetal:01,Eisensteinetal:05}. 
LRGs are passively-evolving,
early-type massive galaxies, and their typical ages are
estimated as $\sim 5~$Gyrs
\citep{Kauffmann:96,Wakeetal:06,Masjedietal:08,CarsonNichol:10}. This
implies that LRGs, at least a majority of the stellar components, were
formed at $z\simgt 1$ \citep{Masjedietal:08}. Motivated by this fact, we
here propose a simplest abundance-matching method for connecting LRGs to
dark matter distribution in large-scale structure as follows.

Our method rests on an assumption that progenitor halos for LRG-host
subhalos 
today
are formed at $z=2$, which is closer to the peak redshift of cosmic star
formation rate \citep{HopkinsBeacom06}.  Our choice of $z=2$ is just the
first attempt, and a formation redshift can be further explored so as to
have a better agreement with the SDSS measurements (see
Section~\ref{sec:conclusion} and Appendix~\ref{app:parameters}
for a further discussion).
(1) We select halos from the simulation output at $z=2$ as 
 candidates of the progenitor halos (hereafter sometimes 
called
$z2$-halo). 
In doing this, we select the $z2$-halos
 in
descending order of their masses (from more massive to less massive)
until the comoving number density becomes close to that of SDSS LRGs at
$z=0.3$, which we set to $\bar{n}_{\rm LRG}=10^{-4}~h^3{\rm
Mpc}^{-3}$. 
{More precisely,}
we need to identify more halos having the number density of 
$\simeq1.3\times10^{-4}~h^3{\rm Mpc}^{-3}$ at least,
because about 30\% of 
$z2$-halos, preferentially in massive halo regions at $z=0.3$,
experience mergers from $z=2$ to $z=0.3$ for the assumed $\Lambda$CDM
model (see below for details).  
(2) We trace the 30\%
innermost particles of each $z2$-halo particles
to lower-redshift simulation outputs until $z=0.3$, where
the
innermost particles are 
considered as ``LRG star'' particles and
defined by particles within
 a spherical boundary around the mass peak of each $z2$-halo 
(see Figure~\ref{fig:snapshot}). 
(3) We perform a matching of the star particles of
each $z2$-halo
to central and satellite subhalos  at $z=0.3$
(hereafter $z0.3$-subhalo). If more
than 50\% of the 
{star}
particles are contained in a
$z0.3$-subhalo, we define the subhalo as a subhalo hosting LRG
 at $z=0.3$.
(4)
We repeat this procedure in descending order of masses of 
$z2$-halos until the comoving
 number density of the matched $z0.3$-subhalos (LRG-host subhalos)
is closest to 
the target value, $\bar{n}_{\rm LRG}=10^{-4}~h^3{\rm Mpc}^{-3}$.
The minimum mass of LRG-progenitor halos at $z=2$ is about $6\times
10^{12}~\hMsun$ for the L1000 run 
(which contains about 90 $N$-body particles for the).

However, 
we need to {\rm a priori} determine some model
parameters
before implementing to the simulation halo/subhalo catalogs: 
the formation redshift of LRG-progenitor halos, $z_{\rm
form}=2$, and the fractions ``30\%'' or ``50\%'' for the star particles
or the matching particles, respectively. Rather than exploring different
combinations of the model parameters to have a better fit to the SDSS
measurements, we will below study the ability of our mock catalog to
predict various statistical quantities of LRGs, by employing our
fiducial choices of the 
parameters ($z_{\rm form}=2$, $f_{\rm star}=0.3$ and $f_{\rm match}=0.5$). 
In Appendix~\ref{app:parameters}, we study how variations of
the model parameters change the mock catalog. 
 The brief summary of the results is the change of each parameter
 affects only the
 small-scale clustering signals, which are sensitive to the fraction of
 satellite LRGs, and does not largely change the clustering signals at
 large scales in the two-halo regime.

In our method,
central and satellite subhalos are populated with 
LRG galaxies under a {{\em single}} criterion:  if a subhalo at $z=0.3$ is a
descendant of the $z2$-halo, the subhalo is included in the matched sample.
On the other hand, 
the standard abundance matching method often
uses different mass proxies
of central and satellite subhalos  when
matching subhalos 
to the target galaxies (in the order of their
stellar masses or luminosities).
 For instance,
the mass of a central subhalo is assigned by
a maximum circular velocity of the bounded $N$-body particles, computed
from the output
at the target redshift ($z=0.3$ in the LRG case), while the mass of a
satellite subhalo is assigned by the 
maximum circular velocity from the simulation output at 
the ``accretion'' epoch before
the subhalo started to accrete onto the main host
halo \citep{Conroyetal:06}, which allows one to estimate the 
mass of  each satellite subhalo before being affected by 
the tidal stripping during penetrating the main halo.
Thus 
the standard abundance matching method is computationally more expensive
in a sense that it requires 
many simulation outputs at different redshifts 
in order to trace the accretion/assembly history of each subhalo. 
To be fair, we below compare
our method with the standard abundance matching
method for some statistical quantities of LRGs. 

Some of the LRG-host halos at $z=0.3$, especially massive halos, contain
multiple LRG-subhalos in our mock catalog (see the example in the
lower panel of 
Figure~\ref{fig:snapshot}). We often call such systems 
``multiple-LRG systems'' in the following discussion \citep[also
see][]{ReidSpergel09,Hikage12b}. We 
refer to 
the LRG-host halos, which
host only one LRG inside, as ``single-LRG systems''.  
The average halo masses for the single- and multiple-LRG systems
are found from the L1000 run to be $\bar M_{\rm vir}=4.8\times 10^{13}$ and 
$1.5\times10^{14}~\hMsun$, respectively. 
The fraction
of the multiple-LRG systems among all the LRG-host halos is about 8\%
in the L1000 run.
Because we assumed that 
 most stars of each LRG are formed
until $z=2$ and the total stellar mass scales with 
the mass of $z2$-halo, 
we define the brightest LRG (BLRG)
in each multiple-LRG system by
the LRG-subhalo that corresponds to 
 the most massive
$z2$-halo among all the progenitor $z2$-halos in the system,
while we
call 
the smallest $z2$-halo 
the faintest LRG (FLRG). Note that we
also
refer to an LRG in a single-LRG system as BLRG.
A BLRG in a single-LRG system 
is not necessarily
a central galaxy in the parent halo
at $z=0.3$ (in other words, the central subhalo does not correspond to
any LRG-progenitor halo at $z=2$).
Similarly, 
a central LRG in a multiple-LRG system
is not necessarily a BLRG, i.e.
the most massive $z2$-halo, although
the central subhalo
is the most
massive subhalo in the parent halo by definition.

Table~\ref{tab:lrgs} summarizes properties of the LRG-host halos 
computed from the L1000 and L300 mock catalogs. 
To estimate statistical uncertainties of each quantity, we divided
the L1000 catalog into 27 sub-volumes (the side length of each
sub-volume is $333~\hMpc$) and computed the mean and rms of the
quantity\footnote{Note that, when computing the mean value
from the 27 sub-volume catalogs, 
we did not use any weight, e.g. by halo
mass. For this reason, 
the relation,
$\bar{M}_{\rm vir}=4\pi\bar{\rho}_{m0}
\Delta_{\rm vir}\bar{R}_{\rm vir}^3/3$, does not hold 
for the mean halo mass and the mean virial radius for
the LRG-host halos in Table~\ref{tab:lrgs}.}. 
Hence the error quoted for each entry of the L1000 run
corresponds to the sample variance scatter
for a volume of $[333~\hMpc]^3$. 
The L1000 result with the error bar can be compared with the L300
result, 
because of
the similar volumes of the sub-divided L1000 catalog
and L300 run ($333^3$ and $300^3$~$[\hMpc]^3$, respectively). 
The L1000 and L300 results agree
with each other to within $2\sigma$ for 
the quantities
except for the fraction of satellite LRGs for all the LRG-host halos.
The disagreement for the satellite LRG fraction 
is probably due to the
numerical resolution, because the L1000 simulation may miss 
some less-massive LRG progenitor-halos 
at
$z=2$, which are identified in the L300 run, 
 due to lack of the numerical resolution and such
small $z=2$-halos 
preferentially become satellite LRGs at $z=0.3$ (also see below and
Appendix~\ref{app:parameters}).

In Table~\ref{tab:lrgs}, we also compare the mock results with
the measurement results
from the SDSS DR7 LRG
 catalog in \cite{Hikage12b}. 
The SDSS results were derived using
the different clustering measurements,
 the LRG-galaxy lensing, the LRG redshift-space power spectrum, and the
 LRG-photometric galaxy cross-correlation to constrain the properties of
 LRG-host halos. To be conservative, we here quote the measurement
 result that has largest uncertainties among the three measurements.
The table shows  
that the mock catalog fairly well reproduces the SDSS results within the
  error bars.
Although one may notice sizable disagreement for the single-LRG systems,
  especially for the fraction of halos hosting satellite LRGs
 ($f_{\rm sat-LRG}$) or the
  fraction of central BLRGs ($q_{\rm cen}^{\rm BLRG}$), the SDSS
  measurements are not yet reliable for the single-LRG systems, as
  reflected by the large error bars and stressed in \cite{Hikage12b}.
Hence, this requires a further careful study.

Figure~\ref{fig:snapshot} shows snapshots of the $N$-body particle
distribution in the L1000 run outputs at different redshifts, for the
regions where multiple- or single-LRG systems are formed at
$z=0.3$. The figure 
illustrates
how each LRG-progenitor halo is
defined at $z=2$, how the innermost particles are assigned as
``star'' particles, and how the star particles
are traced to lower
redshifts and how LRG-progenitor halos merge with each other and become
to reside in
central and satellite
 subhalos at the final redshift $z=0.3$. Our method allows us
to directly include 
the merging and assembly histories of LRG-progenitor
halos. Although the number density of LRG-host subhalos is set to the
density of LRGs as we described above, the figure shows that more
LRG-progenitor halos or subhalos survive at higher redshift than
at $z=0.3$. Hence our method has a capability to study what kinds of
halos or subhalos at higher redshift are progenitors for the SDSS LRGs 
(see Section~\ref{sec:conclusion} for a further
discussion). 

Figure~\ref{fig:mz2-mz0.3} shows how each LRG-progenitor halo at $z=2$
loses or gains its mass due to mass accretion,
 merger and/or tidal stripping when it
becomes an LRG-host subhalo at $z=0.3$, computed using the catalogs of
halos and subhalos in the $z=2$ and $z=0.3 $ outputs of L1000 run.
Note that the halo mass shown in the $x$-axis, $M_{\rm FoF}(z=2)$, is the FoF
mass, the sum of FoF particles in each halo region at $z=2$. 
First, the figure shows that we need to select the LRG-progenitor halos
at $z=2$ down to a mass scale of about $6\times 10^{12}~\hMsun$. 
Some subhalos for satellite LRGs lose their
masses due to tidal stripping as implied in Figure~\ref{fig:snapshot},
while subhalos for central LRGs gain their masses due to
mass accretion
and/or merger. Comparing the left and right panels manifests that multiple-LRG
systems tend to reside in more massive LRG-progenitor halos at $z=2$ and
become more massive LRG-host halos at $z=0.3$, and that the mass difference
between subhalos for central and satellite LRGs is larger in
multiple-LRG systems, implying a larger difference between their
luminosities \citep[see][for a similar discussion]{Hikage12b}.

Thus our method is primarily 
based on the masses of LRG-progenitor halos at
$z=2$ (see Figure~\ref{fig:mz2-mz0.3}) and the connection with central
and satellite subhalos in the parent halos at $z=0.3$.  On the other
hand, LRGs in the SDSS catalog are selected based on the magnitude and
color cuts from the SDSS imaging data (primarily $gri$), and are not
necessarily a stellar-mass-selected sample, although their stellar
masses are believed to have a tight relation with the host halo masses.
Nevertheless, we will show below that the mock catalog perhaps 
surprisingly well
reproduces the different SDSS measurements.

Since LRGs in our mock catalog reside in relatively massive halos at
$z=2$, with masses $M_{\rm FoF}\simgt 6\times 10^{12}~\hMsun$
(Figure~\ref{fig:mz2-mz0.3}), as well as in massive parent halos at
$z=0.3$, our method does not necessarily require 
a high-resolution simulation. A simulation with $1024^3$ particles and 
$1~\hGpc$ size on a side
seems sufficient, which allows for
a relatively fast computation of the $N$-body
simulation as well as
an
accurate estimation of statistical quantities of LRGs. This is not the
case if one wants to work on the abundance matching method for less
massive galaxies or more general types of galaxies
\citep[e.g.,][]{Trujillo-Gomezetal:11,Reddicketal:12,Masakietal:12a}.

\section{Results: comparison with the SDSS LRG measurements}
\label{sec:results}

\subsection{Halo occupation distribution and properties of satellite LRGs}
\label{sec:offset}

\begin{figure}
\begin{center}
\includegraphics[width=8.5cm]{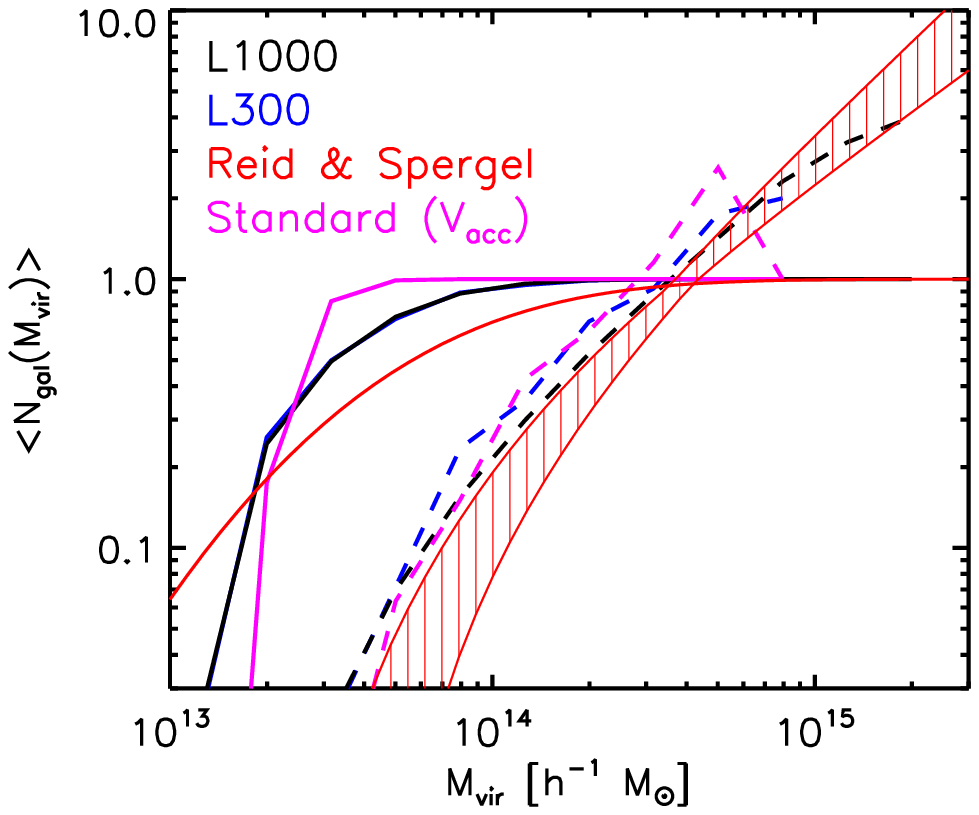}
\caption{The halo occupation distribution (HOD) for LRGs 
  as a function of parent halo mass, measured from 
  our mock catalog. Our mock catalog
  has an assignment of each LRG to central or satellite subhalos in a
  parent halo at $z=0.3$ 
  (see Figure~\ref{fig:snapshot}), thereby allowing us to compute the HODs for central (solid
  curve) and satellite (dashed) LRGs.  The black and blue curves are the
  results from the L1000 and L300 runs, respectively, where the L300
  run is a higher resolution run with a small box size, $300~\hMpc$ 
  (see text for
  details).  The red curves show the SDSS measurements, taken from
  \citet[RS09]{ReidSpergel09}. RS09 fixed the function form of central
  HOD, and then constrained the satellite HOD from the SDSS LRG catalog
  using the Counts-in-Cylinders technique. The hatched region is the
  range allowed by varying each model parameter of the satellite HOD within
  its $1\sigma$ confidence range. 
  The mock catalog  well reproduces the
  SDSS measurements, including the shape of central HOD around the
  cutoff mass scale as well as the slope and amplitude of satellite
  HOD, without employing any free parameter to adjust 
  after the abundance
  matching. 
The magenta lines show the HODs from the LRG mock catalog
 generated using 
 the standard abundance matching method in \citet{Conroyetal:06} (also
 see text for the details).
\label{hod}}
\end{center}
\end{figure}
\begin{figure}
\includegraphics[width=8.5cm]{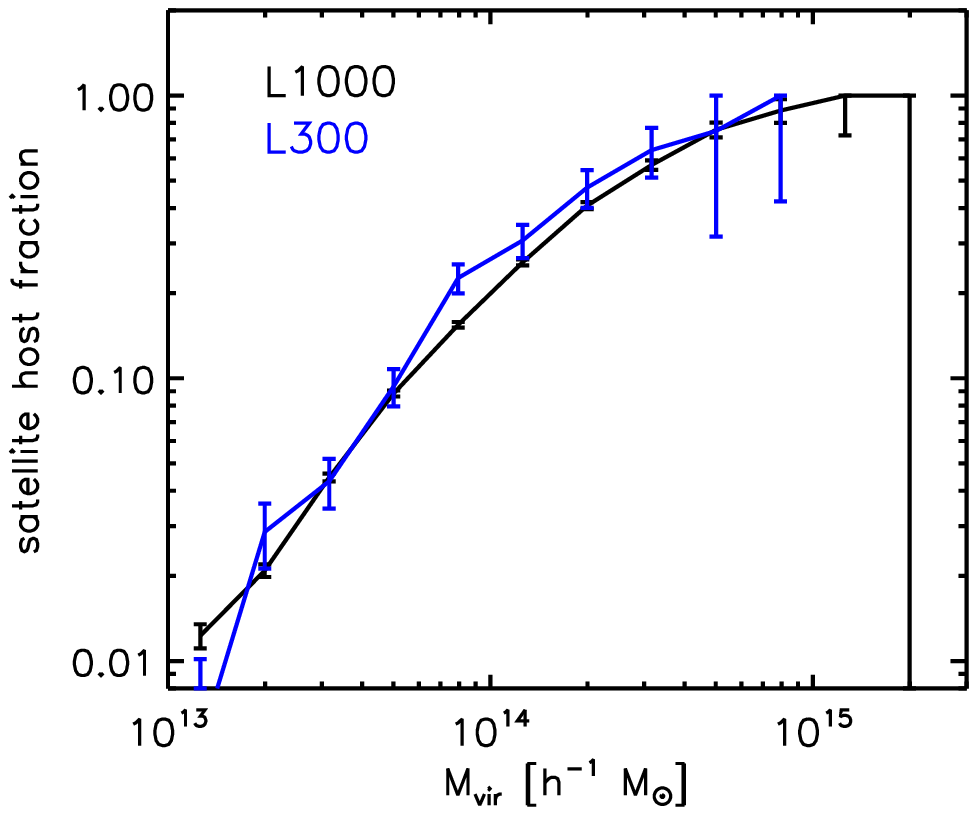}
\caption{The fraction of halos hosting {\em satellite} LRG(s) inside as
 a function of halo mass, computed by using all the LRG-host halos at
 $z=0.3$ in the L1000 and L300 runs.}  \label{satfrac}
\end{figure}
First, we study the halo occupation distribution (HOD) for LRGs in
Figure~\ref{hod}, where the HOD gives the average number of LRGs that 
the halos at $z=0.3$ host as a function of host-halo mass.
Here we consider the HODs
for central and satellite LRGs which reside in central and satellite
subhalos in the LRG-host halos, respectively.
Again we should emphasize that our method does not assume any functional
forms for the HODs, unlike done in the standard HOD method, and rather
allows us to directly compute the HODs from the 
mock catalog.
Even
if LRG-progenitor halos are selected from halos at
$z=2$ by a {\em sharp} mass threshold, our mock catalog naturally
predicts that the central HOD has a smoother shape 
around a minimum halo mass, 
as a result of their
merging and assembly histories from $z=2$ to $z=0.3$.  To be more
precise, the central HOD is 
 smaller than unity ($\ave{N_{\rm cen}}<1$) for host halos 
with $M_{\rm vir}\simlt 10^{14}~\hMsun$, meaning that only
some fraction of the halos host a central LRG. 
On the other hand, 
most
of massive halos host at least one LRG and can host multiple LRGs
inside. 
Conversely, 
the fraction of
massive halos, which do {\em not} host {\em any} LRG,
is 1.3\% for halos with masses $M_{\rm vir} \ge 1
\times 10^{14}~\hMsun$, while {\em all} halos with $M_{\rm
vir}\ge 2\times 10^{14}~\hMsun$ have at least one LRG inside.

To test validity of our mock catalog, we 
compare the HODs with the SDSS measurement
 in \citet[hereafter RS09]{ReidSpergel09}, 
where the HOD was constrained by
using the Counts-in-Cylinders (CiC) method for identifying multiple LRG systems from 
the SDSS DR7 LRG catalog with the aid of halo catalogs in $N$-body simulations.
Although RS09 employed the slightly different cosmological model
and redshift ($z=0.2$) from ours ($z=0.3$),
we
employed the same best-fit parameters in RS09 to compute the LRG HOD 
 for this figure. To be more precise, 
due to limited constraints from the SDSS LRG catalog,
especially for low-mass host-halos, 
RS09 assumed the fixed form for
the central HOD:
\begin{equation}
\ave{N_{\rm cen}(M)}=\frac{1}{2}\left[
1+{\rm erf}\left(\frac{\log M-\log M_{\rm min}}{\sigma_{\log M}}\right)
\right],
\end{equation}
with $M_{\rm min}=5.7\times 10^{13}~\hMsun$ and $\sigma_{\log M}=0.7$,
in order to obtain meaningful constraints on
the satellite HOD. The central HOD
for low-mass host-halos is difficult to constrain, because low-mass host-halos of LRGs are
observationally difficult to identify. 
Therefore, we do not think that
the difference for the central HODs is significant, and needs to be
further carefully studied. 

On the other hand, the  satellite HOD in RS09
is almost perfectly recovered by our mock catalog, where RS09
assumed the functional form for the satellite HOD to be given by
$\ave{N_{\rm
sat}(M)}=\ave{N_{\rm cen}(M)}
[(M-M_{\rm cut})/M_{1}]^{\alpha}$  
and then constrained the
parameters $(M_{\rm cut},M_1, \alpha)$ from the SDSS LRG catalog. The
hatched region is the range at each host-halo mass bin that
is allowed by varying the model parameters within the
$1\sigma$ confidence regions.
Our
results confirm that parent halos of $\sim 10^{15}~\hMsun$ have up to
several LRGs inside, as first pointed out in RS09. The
L300 result, the simulation result with higher spatial resolution, gives
similar results to the L1000 results, showing that the numerical
resolution is not an issue in studying the satellite HOD.  Even though SDSS LRGs
are selected by the magnitude and color cut, not by their masses,
our method seems to capture the origin of SDSS LRGs;
mass-selected halos at $z\sim 2$ are main progenitors of LRGs, and their
subsequent assembly and merging histories determine where LRGs are
distributed within the host halos at lower redshift.

Furthermore, to be comprehensive, we also compare
our method with the standard abundance matching method in 
\citet{Conroyetal:06}.
In this method, the mass proxy of each subhalo 
 is assigned by 
 the maximum circular velocity 
$V_{\rm cir}$ computed from the member $N$-body particles.
More specifically, 
the central subhalo mass is assigned by
$V_{\rm cir}$ 
at the LRG redshift $z=0.3$,
while the satellite subhalo mass
is estimated by $V_{\rm cir}$ from the simulation output
at its ``accretion'' epoch when the subhalo started to accrete 
onto the parent halo at $z=0.3$ (more exactly, the circular velocity
 is estimated
from
the last output when
the ``subhalo''
was identified as an ``isolated'' halo before the accretion)
\citep[see also][]{Masakietal:12a}. This prescription for
satellite subhalos allows for a better assignment of 
the subhalo mass so that it avoids the effect by 
tidal stripping during accreting onto the parent halo.
We use the L300 run outputs at 44 different redshifts from 
$z=10$ to trace the merging and assembly history of each subhalo
till $z=0.3$.
Then, assuming that the stellar masses of LRGs trace the subhalo masses, 
we match the $z=0.3$ subhalos to LRGs 
in descending order of the mass 
proxies $(V_{\rm cir})$ until the number density is closest to the target
value, $\bar{n}_{\rm LRG}=10^{-4}~(h~{\rm Mpc}^{-1})^3$.
The curves labeled as ``Standard ($V_{\rm acc}$)'' show the central and
satellite HODs measured from the mock catalog of 
the $V_{\rm cir}$-based abundance matching method. 
The satellite HOD is in a nice agreement with our method, while 
 the central HOD  
from the abundance matching method displays a sharper cut-off
than in our method. Again we do not yet know the genuine cut-off feature of
the central HOD due to lack of the measurement constraints.  
We will below further compare our method with the
abundance matching method for other statistical quantities of LRGs.

One motivation of this paper is to understand the physics of the
nonlinear redshift-space distortion, i.e. the Finger-of-God (FoG) effect,
in the redshift-space power spectrum of LRGs.
The FoG effect
is caused by internal motion of satellite LRG(s) in LRG-host halos
\citep{Hikage12,Hikage12b}.  In the following, we study several
quantities relevant for the FoG effect; the fraction of satellite LRGs,
the radial profile of satellite LRGs inside the parent halos and the
internal velocities of satellite LRGs \cite[see][for details of the
theoretical modeling]{Hikage12}.

Figure~\ref{satfrac} shows how much fraction of LRG-host halos at
$z=0.3$ host {\em satellite} LRG(s) inside, as a function of the halo
mass.
Note that we excluded halos that do not host any LRG in this statistics,
but included the single-LRG
systems hosting one LRG as a {\em satellite} galaxy 
when computing the numerator of the fraction (in this case, the
central subhalo of the parent halo does not correspond to  any 
LRG-progenitor halo at $z=2$).
The error bars around the solid curve are Poisson
errors, estimated using the number of halos in each mass bin.  The
figure shows that more massive halos have a higher probability to host
satellite LRG(s). About 20\% of parent halos with $M_{\rm vir}\simeq
10^{14}~\hMsun$ host satellite LRG(s).

\begin{figure}
\includegraphics[width=8.5cm]{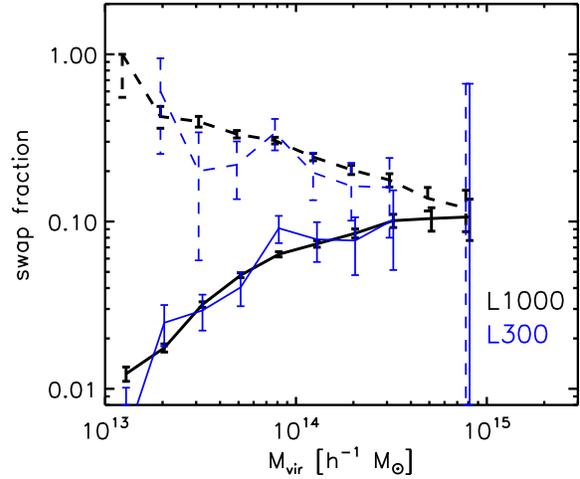}
\caption{The solid curves show the fraction of the parent halos hosting
 the brightest LRG (BLRG) as a {\em satellite} galaxy, among all the LRG-host
 halos.
 Here the BLRG is the most
 massive LRG-progenitor halo at $z=2$ among all the progenitor halos
 which become to reside in the same LRG-host halo at $z=0.3$.  The dashed curves
 are the similar fraction of LRG-host halos with satellite BLRG, but
 computed using only the
 multiple-LRG systems.  The error bars are computed from the number of
 halos in each mass bin assuming Poisson statistics.}  \label{blrg_sat}
\end{figure}
\begin{figure}
\includegraphics[width=8.5cm]{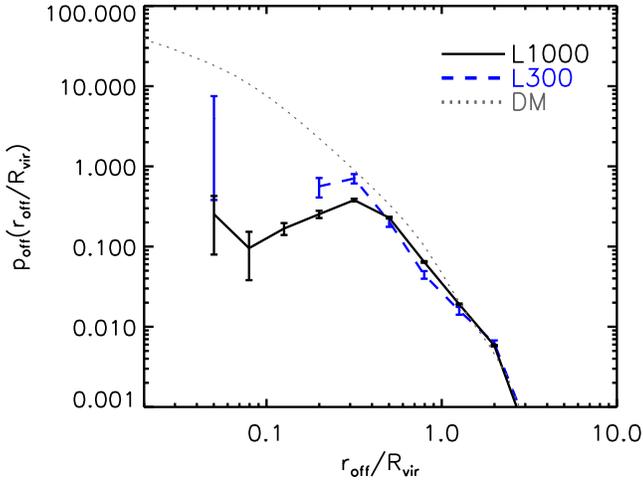}
\caption{The average radial profile of satellite LRG host subhalos, obtained by
 stacking the positions of satellite LRGs in all the LRG-host halos with
 satellite LRG(s), as a function of radius relative to the virial radius
 of each parent halo. The mean mass of the LRG-halos used in this
 calculation is $M_{\rm vir}\simeq 1.31$ or 
$1.24\times 10^{14}~\hMsun$ for the
 L1000 or L300 runs, while the mean virial radius is
 $R_{\rm vir}\simeq 1.07$ or $1.06~\hMpc$, respectively. For
 comparison, the upper dotted curve shows the profile of dark matter
 averaged for the same host halos with an arbitrary amplitude.
 The error bars at each radial
 bin are estimated by first dividing LRG-host halos into 27 subsamples
 (27 subvolumes) and then computing variance of the number of satellite
 LRGs at the radial bin.
The typical off-center radius
 for satellite LRGs appears to be about $400~\hkpc$. } \label{poff}
\end{figure}
\begin{figure}
\includegraphics[width=8.5cm]{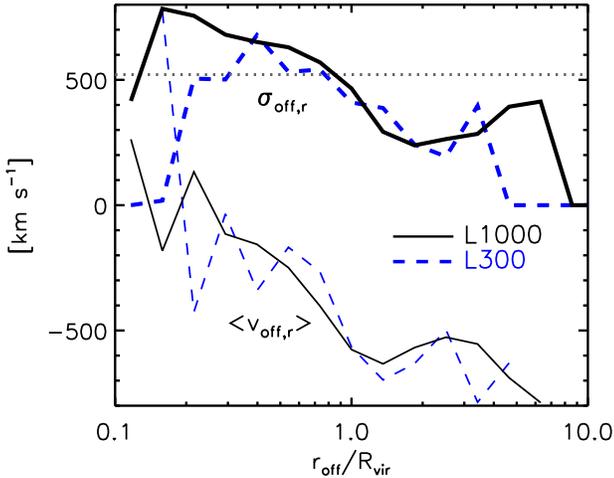}
\caption{The average radial velocity of satellite LRGs, $\ave{v_{{\rm
 off},r}}$, with respect to the halo center in each LRG-host halo,
 computed by using all the LRG-host halos with satellite LRG(s) as in
 the previous figure. The negative $\ave{v_{{\rm off},r}}$ means a
 coherent infall towards the halo center. The upper curves show the
 average radial velocity dispersion around the coherent infall,
 $\sigma_{{\rm off},r}$. 
 For the comparison, the dotted line shows the
 average velocity dispersion expected from virial theorem, $\sigma_{\rm
 vir}=\sqrt{GM_{\rm vir}/2R_{\rm vir}}$.
 The combination of
 $\ave{v_{{\rm off},r}}$ and $\sigma_{{\rm off},r}$ implies that
 satellite LRG(s) sink towards the halo center due to  dynamical
 friction, and then have an oscillating motion around the halo center
 with the velocity dispersion of $\simeq500~{\rm km~s^{-1}}$.  } \label{vrad}
\end{figure}

We naively expect that BLRG, the most massive LRG-progenitor halo at
$z=2$ among LRG-progenitor halo(s) 
accreting onto the same parent halo at
$z=0.3$,
becomes a central galaxy.
 The solid curves in 
Figure~\ref{blrg_sat} show the fraction of BLRGs to be 
a satellite galaxy in LRG-host halos at $z=0.3$ as a function of
the halo mass, 
computed using all the LRG-host halos.
For halos with $M_{\rm vir}\simgt 10^{14}~\hMsun$, 
there is up to 10\% probability for its BLRG to be a satellite galaxy. 

The dashed curves are the similar fraction, but 
computed using only the multiple LRG systems. This sample is
intended to compare with the recent result in \cite{Hikage12b} 
(also see Table~\ref{tab:lrgs}).
In this case, the fraction of satellite BLRGs is higher for host
halos with smaller masses, with larger error bars.
This can be explained as follows. 
Most of low-mass host-halos with masses 
$\simlt 10^{14}~\hMsun$ are single-LRG systems as can be found from
Figure~\ref{hod}, and only a small number of such halos are multiple-LRG
systems, causing larger Poisson error bars at each mass bin. 
We have found from the simulation outputs that such low-mass halos
of multiple LRG systems 
(mostly the systems with 2 LRGs) tend to display 
a bimodal mass distribution due to 
ongoing
or past major merger, where 
the BLRG and
other (mostly central) LRG tend to have the small mass difference.  
As a result, such low-mass multiple-LRG systems have a higher chance to
host the BLRG as a satellite LRG.
On the other hand, the fraction of halos with satellite BLRG
converge to the solid curve with increasing the host-halo mass, because
most of such massive halos are multiple-LRG systems. For multiple LRG systems 
with mass of $M_{\rm vir}\simeq 10^{14}~\hMsun$, about 30\% of BLRGs are 
satellite galaxies.

Recently, \cite{Hikage12b} studied the multiple-LRG systems defined from
the SDSS DR7 catalog by applying the CiC technique as well as
 the FoF group finder
method to the distribution of LRGs in redshift space. Then they used the
different correlation measurements, the redshift-space power spectrum,
the LRG-galaxy lensing and the cross-correlation of LRGs with photometric galaxies, 
to study properties of satellite LRGs.  
{From the lensing analysis,} they found that
the multiple-LRG systems has a typical halo mass of {$M_{\rm vir}\simeq
1.5\times 10^{14}~\hMsun$} (with a roughly 10\% statistical
error),
and that {$37\pm21$\%} of BLRGs in the
multiple-LRG systems appear to be satellite galaxies\footnote{Note that \cite{Hikage12b}  used the 
halo mass definition of $M_{180b}$ instead of the virial 
mass $M_{\rm vir}$ in their analysis, where $M_{180b}$ is
defined by the enclosed mass inside which the mean density is
180 times the mean background mass density.}.
Our mock catalog shows a fairly good agreement with the
SDSS results, for the average halo mass and the fraction of satellite
BLRGs (also see Table~\ref{tab:lrgs}). 

In Figure~\ref{poff}, we study the average radial profile of satellite
LRGs. 
In this calculation, we employ only the host halos
containing satellite LRG(s), and estimate the radial profile by
stacking the radial distribution of satellite LRG(s) in units of the
radius relative to the virial radius of each halo.
We use the mass peak of the smooth component as the halo center.
The average profile $p_{\rm off}$ is normalized as
\begin{equation}
  \int dr'~4\pi r'^2p_{\rm off}(r')=1, ~~~
  {\rm with}~~r'=r_{\rm off}/R_{\rm vir},
\end{equation}
where $r_{\rm off}$ is the distance from the density maximum of the 
smooth component.
The average mass of the host halos is $M_{\rm vir}\simeq 1.31\times
10^{14}$ or $1.24\times 10^{14}~\hMsun$ for the L1000 or L300 run,
respectively, while the average virial radius $R_{\rm vir}\simeq
1.07$ or $1.06~\hMpc$ in the comoving unit.  Compared to the dark matter
profile, the radial profile of satellite LRGs clearly displays a
flattened profile. The typical off-center radius, where the profile
starts to be flattened, is found to be about $400~\hkpc$ because 
$R_{\rm
vir}\simeq 1~\hMpc$, which is in a good agreement with the result for the multiple systems in
\cite{Hikage12b}. 
The radial profile also shows a decline at the
smaller radii. 
Thus our result is not consistent with
the assumption often used in a standard HOD method that the radial profile of
member galaxies follows the dark matter profile \citep[see][for the
improved HOD method including a possible variation in the radial profile
of member galaxies]{BerlindWeinberg02}.
However, the L300 run shows no satellite
LRG at small radii $r_{\rm off}/R_{\rm vir}\le 0.1$, except for the
innermost bin. Thus the satellite LRGs at the small radii are mainly
from most massive host-halos, which do not exist in the smaller box
simulation, L300. Although the mock catalogs show a sharp rise at the
innermost bin $r_{\rm off}/R_{\rm vir}\simeq 0.06~(r_{\rm off}\simeq 60~\hkpc)$,
which may indicate merging LRGs to the central subhalo in the less
massive halos, 
the
scatters are large even for the L300 run, so the result is not
significant.
Nevertheless, it is worth mentioning that 
the satellite LRG 
distribution in our mock catalog
seems to show a similar profile to the profile of most
massive subhalos in cluster-scale halos  in \cite{Gaoetal:12}
(see Figures~15 and 16 for the profile).  
These features in the radial profile of massive subhalos may be as a
result of dynamical friction, tidal stripping and merger to the central
subhalo.
However, the L300 and L1000 results show some difference
at the small scales, so 
a further careful study is needed to derive a more robust
conclusion,  by using
high-resolution simulations as well as a larger number of the
realizations.

Figure~\ref{vrad} shows the average radial profile of internal motions
of satellite LRGs in the parent halos, where the bulk motion of each
parent halo (the average velocity of $N$-body particles belonging to the
smooth component of the halo) is subtracted from the velocity of each
LRG-host subhalo. 
We considered only the host halos with satellite LRG(s) as
in Figure~\ref{poff}. The curves, labelled as $\ave{v_{{\rm off}, r}}$,
are the average radial velocities for all the satellite LRGs with respect
to the halo center. The average velocity is negative, reflecting the
coherent infall motion towards the halo center, and the infall velocity
is larger with increasing radius up to the virial radius. The average
radial velocity becomes zero on average at the halo center. These
support that the LRG-host subhalo approaches to the halo center due to
dynamical friction. On the other hand, the curves, labelled as
$\sigma_{{\rm off},r}$, are the average velocity dispersions of satellite
LRGs. The velocity dispersion has greater amplitudes with decreasing the
radius, reaching to $\sigma_{{\rm off},r}\simeq 500~{\rm km~s^{-1}}$.  
{For comparison, the horizontal dotted line shows the average 
virial velocity dispersion, $\sigma_{\rm vir}\equiv \sqrt{GM_{\rm vir}/2R_{\rm vir}}=521~{\rm km~s^{-1}}$,
among the satellite LRG-host halos in the L1000 run.
}
The combination of $\ave{v_{{\rm off},r}}$ and $\sigma_{{\rm off}, r}$
implies that satellite LRGs gradually approach to the halo center due to
dynamical friction and have  an oscillating motion
 around the halo center.  Again
the amplitude of the velocity dispersion, $\sigma_{{\rm off},r}\simeq
500~{\rm km~s^{-1}}$, is in nice agreement with the recent measurement in
\cite{Hikage12b}, where they found the velocity dispersion of about
$500~{\rm km~s^{-1}}$ for satellite LRGs in the multiple-LRG systems by combining
the different correlation measurements from the SDSS DR7 LRGs. 
{In Section~\ref{sec:pk} we will further discuss how satellite
LRG-subhalos affect the redshift-space power spectrum due to the FoG effect.}

\subsection{Projected correlation function}

\begin{figure}
\includegraphics[width=8.5cm]{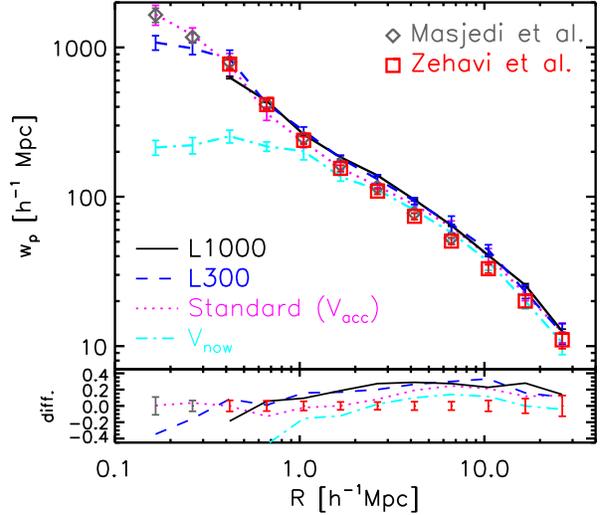}
\caption{{\em Top panel:} Projected auto-correlation function of LRGs,
 $w_p(R)$, as a function of the
projected distance $R$. The solid and dashed curves show the
results from our mock catalogs in the L1000 and L300 runs,
respectively. The error bars are estimated using the measurements
from 8 subdivided volumes of each simulation volume, where 
the error bars are estimated by dividing the standard deviation by
 $\sqrt{8}$. Hence the error bars are the statistical scatters for a
 volume of ($1~\hGpc$)$^3$ or ($300~\hMpc$)$^3$, respectively.
The square and diamond symbols are the correlation functions measured
from the SDSS catalog of LRGs at $z\sim 0.3$, taken from
\citet{Zehavi05} and \citet{Masjedi06}, respectively.
For comparison, the magenta, dotted curve shows 
the result from the standard
 abundance matching method as in Figure~\ref{hod}. Furthermore, we also
 show the prediction obtained if we used the maximum circular velocity
 at the LRG redshift $z=0.3$, instead of the accretion epoch, for
 the satellite subhalos in the abundance matching method (see text for
 the details).
{\em Bottom panel}: The fractional differences of the model predictions
 compared 
to the SDSS measurements.}
\label{wp}
\end{figure}
Next we study the projected auto-correlation function of LRGs,
$w_p(R)$, defined as
\begin{equation}
  w_{p}(R)=2\int_0^{\pi_{\rm max}}
   d\pi~\xi_{gg}(r=\sqrt{\pi^2 +R^2}),
\end{equation}
where 
$R$ is the projected separation between two LRGs in the pairs
used for the correlation measurement in units of the comoving scale, 
$\pi$ is the separation parallel to the line-of-sight
and
$\xi_{gg}(r)$ is the three-dimensional correlation function. 
Following \cite{Zehavi05}, $\pi_{\rm max}$ is set to be $80~\hMpc$.
The projected
correlation function is not affected by the redshift-space 
distortion effect
due to peculiar velocities of LRGs.  

In Figure~\ref{wp}, we compare the
projected correlation function measured from our LRG mock catalog with
the SDSS measurements \citep{Zehavi05,Masjedi06}. In the SDSS
measurements, \cite{Zehavi05} used an LRG sample in the magnitude 
range of
$-23.2<M_g<-21.2$ and with the mean redshift $\ave{z}\simeq0.3$.
\cite{Masjedi06} used the same sample to extend the measurement down
to very small scale, below $R=500~\hkpc$, by taking into account
various observational effects such as the fiber collision. 
Note that the cosmological model
employed in the measurement is slightly different from the model we
assumed for our simulations. 
The figure shows that our mock catalog remarkably 
well reproduces the projected
correlation function of LRGs, to within $30\%$ accuracy 
in the amplitude, 
over a wide range of separation radii, which 
arise from correlations
between LRGs within the same host halo and in different host halos, the
so-called one- and two-halo regimes, 
respectively\footnote{Note that the error bars for the mock catalogs 
are estimated in a different way from those in Table~\ref{tab:lrgs}, and
the error bars in Figures~\ref{wp} -- \ref{fig:dsigma_multi} are the
statistical scatter for a volume of ($1000~\hGpc$)$^3$ or
($300~\hMpc$)$^3$.}.
Comparing the results for the L1000 and L300 runs reveals that the
correlation function for L1000 has a smaller amplitude 
at $R<0.7~\hMpc$
than that for
L300. 
Thus the L1000
run implies a systematic error due to the lack of numerical resolution
at the small scales. 
The L300 result shows a better agreement with the SDSS measurement
in \cite{Masjedi06}. The small-scale clustering arises mostly
from correlation between LRGs in the same multiple-LRG system,
so that numerical
resolution seems important to resolve these small subhalos (also see
below for a further discussion). 

As in Figure~\ref{hod}, the dotted curve
gives the result from the standard
abundance matching method, which shows almost similar-level agreement
with the SDSS measurements to our method. 
Thus, since the abundance matching method rests on the higher-resolution
L300 outputs at 44 different redshifts (in our case), 
our method can provide a much computationally-cheap, alternative
approach to making a mock catalog of LRGs. 
Furthermore, for comparison, 
the dot-dashed curve shows the correlation function, if the abundance matching
is done by using the maximum circular velocity at the LRG redshift
($z=0.3$) for  each satellite subhalo
 as its mass proxy, instead of the velocity
at the accretion epoch.  
The result shows a significant discrepancy with the SDSS measurements or
our method and the standard abundance matching method,
especially at small radii. 
The disagreement means that 
the circular velocity at $z=0.3$ is not a good mass proxy for
satellite subhalos when matching the subhalos to LRGs, because it misses
satellite subhalos in the multiple-LRG systems. To be more precise, 
 mass (circular velocity) of each satellite subhalo
tends to be 
underestimated due to the tidal stripping, then tends to be not selected
by the abundance matching, 
 and instead other isolated, less-massive
halos tend to be selected. This reduces the clustering signals at small
scale due to less contribution from satellite subhalos and also reduces
the clustering signal at large scales due to a smaller bias for such
low-mass halos. Thus detailed features of the correlation functions at
different scales are sensitive to the contribution of satellite LRGs as
well as the low-mass threshold of central HOD in Figure~\ref{hod}
(also see Appendix~\ref{app:parameters}).
Note that an explicit
implementation of the abundance matching method to LRGs is the first
time, and the result in Figure~\ref{wp} highlights the importance of
proper assignment of subhalo masses in the abundance matching method.

\subsection{LRG-galaxy weak lensing}

\begin{figure}
\includegraphics[width=8.5cm]{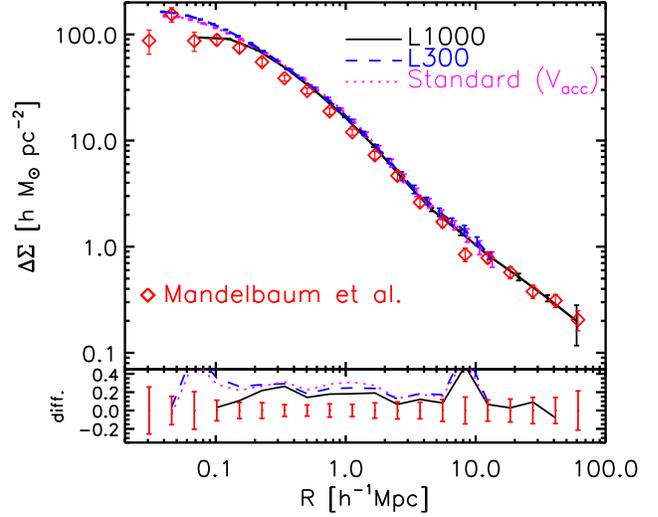}
\caption{
  {\em Top Panel}: The average surface mass density profile around LRGs, which is an
  observable of the LRG-galaxy weak lensing. The solid and dashed curves are the
  results of our mock catalog, obtained by stacking  $N$-body particles
 around  all the LRG-host subhalos in the L1000
  and L300 runs, respectively. 
  The error bars are estimated using the measurements from 27 subsamples of LRG-host subhalos.
  The data with error bars show the SDSS measurements in
  \citet{Mandelbaumetal:12}. 
As in Figures~\ref{hod} and \ref{wp}, we also show the result
 obtained from the standard abundance matching method (dotted curve). 
{\em Bottom panel}: The fractional differences of the model predictions
 compared to the measurement.
  \label{fig:dsigma}}
\end{figure}

Correlating the positions of LRGs with shapes of background galaxies,
the so-called LRG-galaxy weak lensing, is a powerful means of probing
the average dark matter distribution around the LRGs
\citep{Mandelbaumetal:06b,Mandelbaumetal:12}. The LRG-galaxy lensing
measures the radial profile of differential surface mass density defined
as
\begin{equation}
\Delta\Sigma(R)=\bar{\Sigma}(<R)-\Sigma(R).
\label{eq:dsigma}
\end{equation}
The profile $\Sigma(R)$ is the average surface mass density around the
LRGs defined as
\begin{equation}
\Sigma(R)=\bar{\rho}_{m0}\int d\pi[1+\xi_{gm}(r=\sqrt{\pi^2+R^2})],
\end{equation}
where $\bar{\rho}_{m0}$ is the mean
background mass density today, and $\xi_{gm}(r)$ is the
three-dimensional cross-correlation between LRGs and the surrounding
matter. In Eq.~(\ref{eq:dsigma}), $\bar{\Sigma}(<R)$ is the
surface mass density averaged within a circular aperture of a radius
$R$. Our use of the mean mass density today ($\bar{\rho}_{m0}$)
 is due to our use of the comoving units. 

Figure~\ref{fig:dsigma} shows that the average mass profile measured
for all LRGs in the mock catalog is in good agreement with the 
SDSS measurement in \citet{Mandelbaumetal:12}, to within $30\%$ level in
the amplitude.
 Note that, to
obtain the average mass profile from our mock catalog, we stacked all
$N$-body particles around all the LRG-host subhalo in the simulation,
including the particles outside dark matter halos.
The lensing
signal at the radii smaller than about $1~\hMpc$ arises from the mass
distribution within the same halo, while the signal at the larger scale
arises from the mass distribution surrounding the host halos --
 the one- and two-halo terms, respectively \citep[e.g. see][]{OguriTakada:11}. 
The mock catalog
well reproduces both the signals of different scales.
The average halo mass inferred from the SDSS measurement is 
$\bar{M}_{\rm vir}\simeq 4.1\times 10^{13}~\hMsun$ 
\citep{Hikage12b} (see also Table~\ref{tab:lrgs}). 
Furthermore, 
the standard abundance matching method shows a similar-level agreement
with the SDSS measurement, similarly to Figure~\ref{wp}.

\begin{figure*}
\includegraphics[width=17.cm]{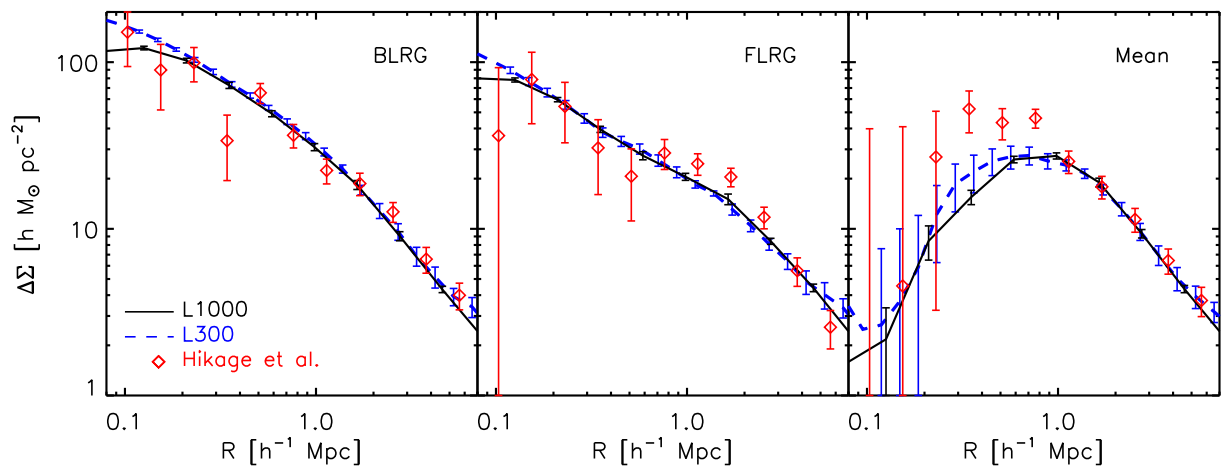}
\caption{The average surface mass profiles for the multiple-LRG systems. The
 different panels show the results obtained by taking the
 different centers in each multiple-LRG halo; the brightest LRG
 (BLRG), the faintest LRG (FLRG) and the center-of-mass of different
 LRGs or the arithmetic mean positions of member LRGs
 (Mean) in the left, middle and right panels, respectively. The data
 with error bars show the SDSS measurements for the multiple-LRG systems
 in \citet{Hikage12b}.
 \label{fig:dsigma_multi}}
\end{figure*}
\cite{Hikage12b} also used the SDSS LRG catalog to study the weak
lensing for the multiple-LRG systems.  When making the lensing
measurements, they used three different proxies for the halo center of
each multiple-LRG system, the BLRG, FLRG and the arithmetic mean
position of member LRGs
(hereafter ``Mean'').  By comparing the lensing signals for the
different centers, they constrained the average radial profiles of
satellite BLRGs and FLRGs, 
finding about $400~\hkpc$ for a typical offset radius from the true center. 
Figure~\ref{fig:dsigma_multi} shows that the mock catalog
predictions are in remarkably good agreement with the SDSS measurements
for the different centers. Since these lensing signals are from the
exactly same catalog
 of the multiple-LRG systems, the differences
between the different measurements should be due to the off-centering
effects of the chosen centers.
  As nicely shown in
\cite{Hikage12b}, the lensing signals for the BLRG and FLRG centers can
be well explained by a mixture of the central and satellite BLRGs or
FLRGs 
in the
sample. The lensing signals for the FLRG center have smaller amplitudes
due to the larger dilution effect because of a larger fraction of
satellite (off-centered) FLRGs than in the BLRG centers.
On the other hand, the Mean center does not have any galaxy (subhalo) at
its position, and therefore the Mean center always has an off-centering
effect from the true center in each LRG system.  This causes decreasing
powers of the lensing signal at the smaller radii than 
the typical off-center radius.  
The lensing signals at some radii for the FLRG and Mean centers show
some discrepancy from the mock catalog, but we do not think that the
disagreement is significant. 
The average masses inferred from the SDSS measurement and the mock
catalog 
for the multiple-LRG halos 
 agree within about 30\%; 
{$\bar{M}_{\rm vir}\simeq 1.46$} or $1.52\times 10^{14}~\hMsun$,
respectively (also see Table~\ref{tab:lrgs}).  

As can be shown in Figures~\ref{wp}, \ref{fig:dsigma} and \ref{fig:dsigma_multi}, our mock
catalog of LRGs well reproduces both the SDSS measurements for the
auto-correlation function of LRGs and the LRG-galaxy weak lensing
simultaneously. As
recently discussed in \cite{NeisteinKhochfar:12} \citep[also
see][]{Neisteinetal:11}, the abundance matching method
has a difficulty to
reproduce these measurements with the same model, although they
considered the spectroscopic sample of SDSS galaxies, rather than
focused on LRGs. Thus the agreements of our mock catalog show a
capability of our method to predict different statistical quantities of
LRGs by self-consistently modeling, rather than assuming, the fractions
of satellite LRGs among different halos and the radial distribution of
satellite LRGs in the parent halos
 \citep[also see][for a recent development on the extended abundance
 matching method
based on the similar motivation]{Masakietal:12a}.

\subsection{Redshift-space power spectrum of LRGs}
\label{sec:pk}

\begin{figure*}
\includegraphics[width=8.5cm]{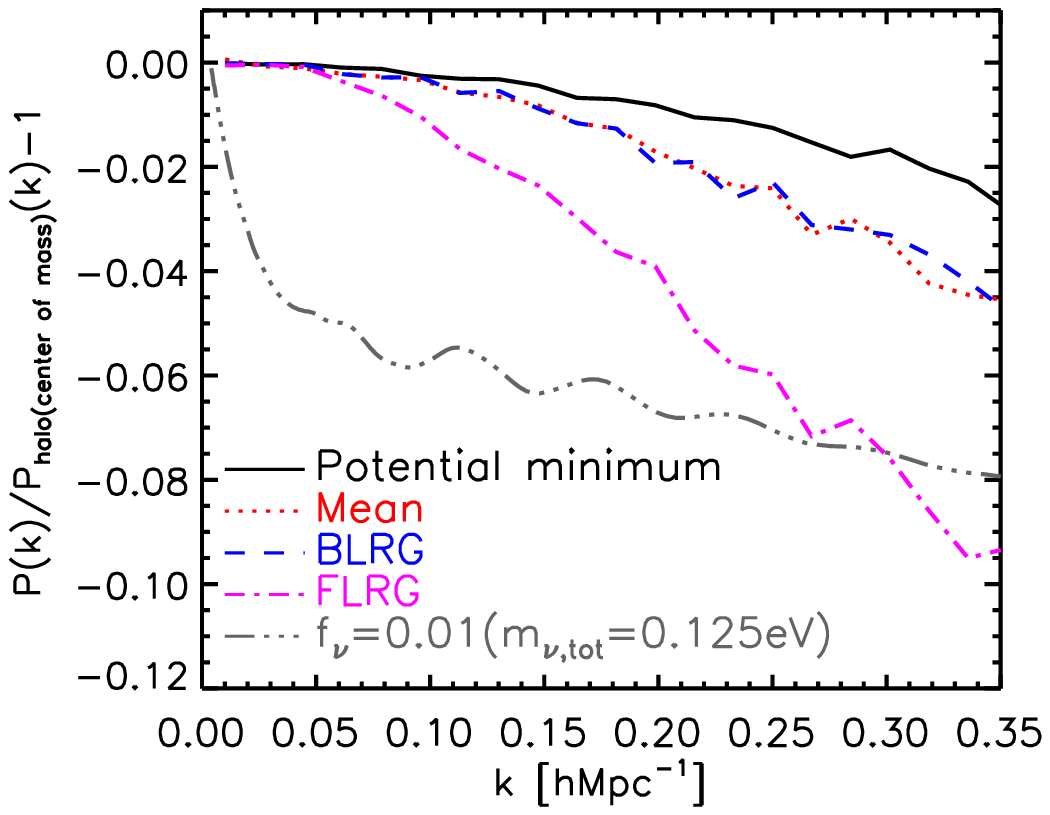}
\includegraphics[width=8.5cm]{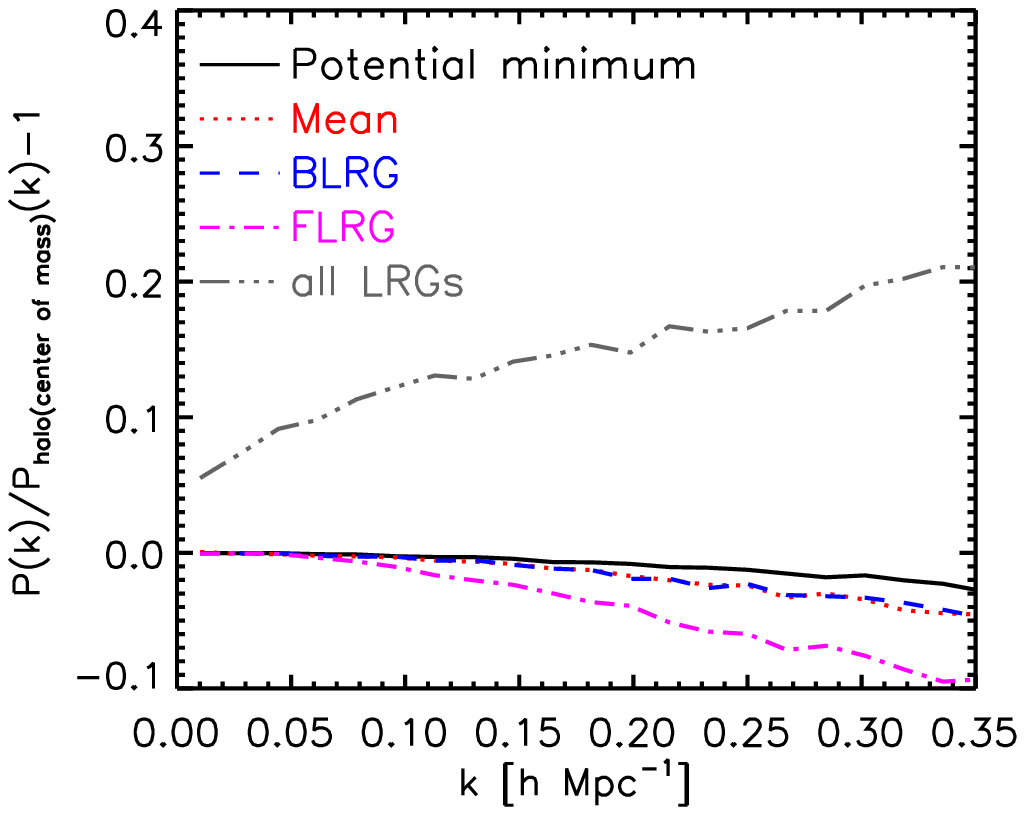}
\caption{The angle-averaged redshift-space power spectra for the LRG-host
halos at $z=0.3$, computed from the L1000 run.  The different curves
 show the fractional differences of the power spectra using different
 proxies of each LRG-host halo position in the power spectrum estimation, 
relative to the power spectrum for the mass center as the halo position.
{\em Left panel}: The dotted, dashed and dot-dashed curves are the
 results when using different halo center proxies for 
 each multiple-LRG system;
the arithmetic mean position of the member LRGs 
in redshift space (Mean), BLRG or FLRG
 as in Figure~\ref{fig:dsigma_multi}. Note
that we use the LRG position 
as the halo center
for each single-LRG system. 
Thus the
differences between the different spectra arise only from the different
positions of multiple-LRG systems in redshift space, to be compared with
 \citet{Hikage12b}. The different power spectra show decreasing amplitudes
with
 increasing wavenumber, which is caused by the nonlinear redshift-space
 distortion,  the so-called Finger-of-God effect, 
due to the internal motions of the chosen halo centers in LRG-host halos.
For comparison, we also show the power spectrum
measured using the potential minimum of each LRG-host halo, 
where the potential minimum is the mass density peak of the
smooth component of the halo that is likely to host the central
 galaxy. For comparison, the three dots-dashed curve shows the effect on the real-space matter power spectrum caused by
 massive neutrinos assuming the total neutrino mass $m_{\nu,{\rm tot}}=0.125~{\rm eV}$.
{\em Right panel}: Similar to the left panel, but the power spectrum
 using all the LRGs is added (the three dots-dashed curve). The power
 spectrum includes contributions from multiple LRGs in the same halo. 
The shot noise contamination due to the different number densities of the LRG-host
 halos and the LRG-host subhalos is properly subtracted to have a fair
 comparison. 
\label{fig:pk}}
\end{figure*}

\begin{figure*}
\begin{center}
 \includegraphics[width=8.5cm]{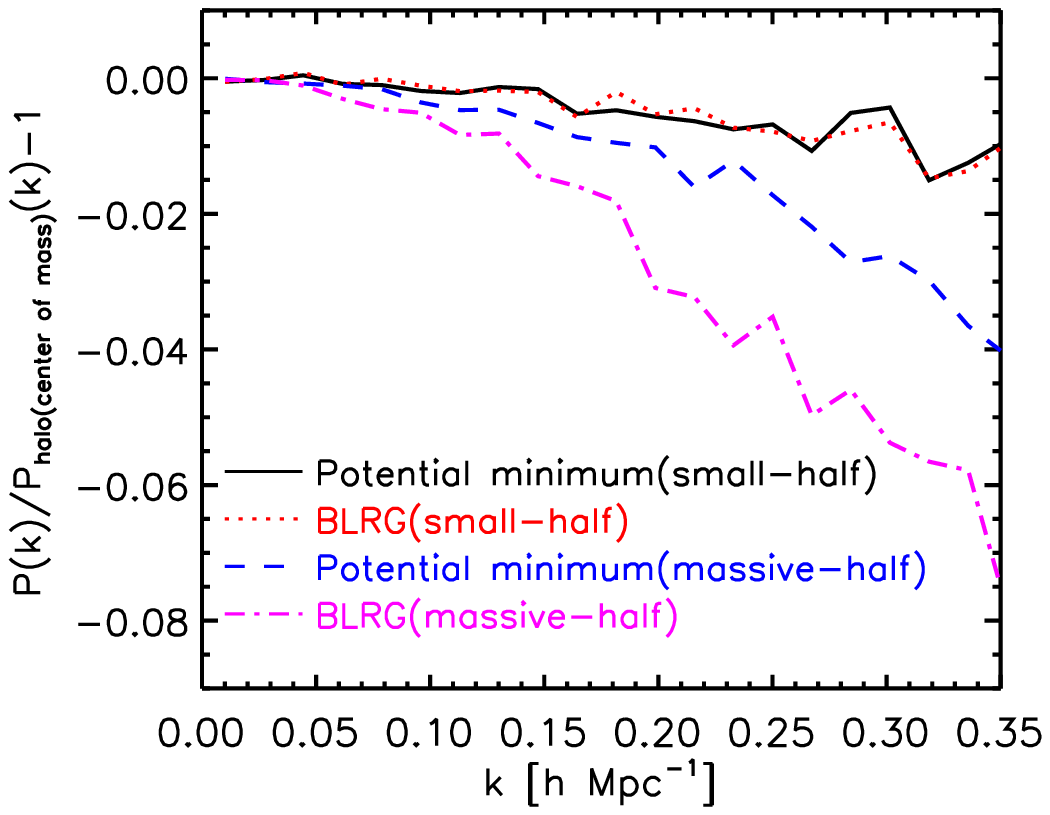}
\includegraphics[width=8.5cm]{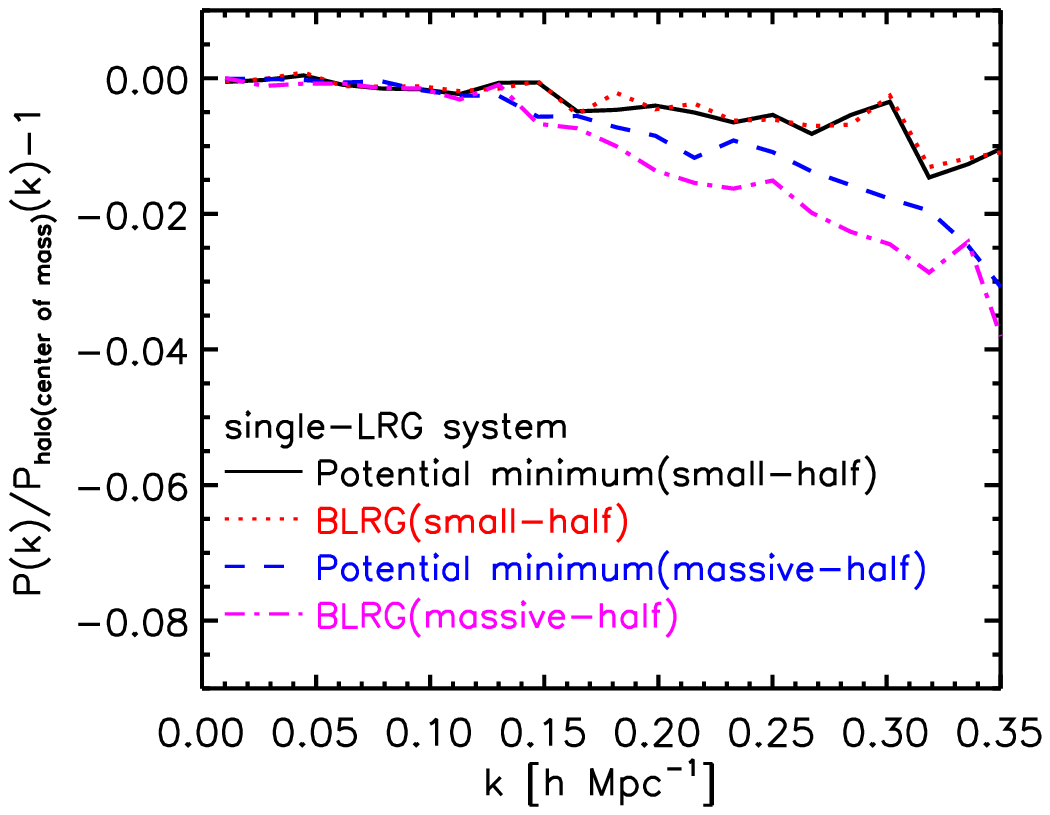}
\caption{Similarly to the previous figure, but for the halved samples of
 LRG-host halos. {\em Left panel}: The LRG-halos are divided into two
 halved samples by the halo masses; one sub-sample is defined by halos
 which have masses smaller than the median 
 mass (``small-half''), while the other is by halos with masses larger
 than the median (``massive-half''). The massive halo sub-sample shows a
 stronger FoG effect.   
{\em Right panel}: Similar plot, but using only the single-LRG
 halos. 
\label{fig:pk_halved}}
\end{center}
\end{figure*}

Another observable we consider is the redshift-space power spectrum of
LRGs. The FoG effect due to internal motion of galaxies is one of
systematic errors to complicate the cosmological interpretation of the
measured power spectrum.  The FoG effect involves complicated physics
inherent in the evolution and assembly processes of galaxies, so is very
difficult to accurately model from the first principles. One
way to reduce the FoG contamination is to remove satellite galaxies from
the region of each multiple-LRG system, and to keep only one galaxy (LRG
in our case), ideally the central galaxy, because the central galaxy is
supposed to be at rest with respect to the parent halo center and does
not suffer from the FoG effect.
For example, \citet{Reidetal:10} developed a useful method
for this purpose; first, reconstruct the distribution of halos from the
measured distribution of LRGs by identifying multiple-LRG systems based
on the 
CiC and FoF
group finder method, and then keep only one
LRG for each multiple-LRG system.  However, the chosen LRG is not
necessarily the central galaxy (more exactly, they used, as the halo
center proxy, the arithmetic 
mean of member LRGs or the center-of-mass of different CiC groups 
without any luminosity or mass weighting), so
there may generally remain a residual FoG contamination in the
measured LRG power spectrum as pointed out in \cite{Hikage12b}.

In the left panel of Figure~\ref{fig:pk}, we study the FoG effect on the
redshift-space power spectrum, caused by the off-centering effect of
LRGs in our mock catalog. Following the method in \cite{Reidetal:10} and
\cite{Hikage12b}, we study the redshift-space power spectrum for
LRG-host halos,
instead of the power spectrum for LRGs. To compute the
power spectrum of halos, we need to specify 
the
halo center in each LRG-host halo.
For single-LRG systems, we use the LRG position as the halo center
proxy. For multiple-LRG systems, we employ different proxies of halo
center for each system as done in Figure~\ref{fig:dsigma_multi} for the
LRG-galaxy lensing; BLRG, FLRG or the arithmetic mean  (Mean), where the
Mean center is computed in redshift space taking into account redshift
space distortion due to peculiar velocities of LRG-subhalos.  The
figure shows the angle-averaged redshift-space power spectra for the
different centers, relative to the power spectrum for the mass center of
each LRG-host halo (the mass center of $N$-body particles of the host
halo).  Note that, for the power spectrum measurement, we used the
exactly same catalog of LRG-host halos, and the different
power spectra differ in the chosen halo center  
of each 
multiple-LRG system.
Hence, the
difference between the different spectra should be from the
off-centering effects of the chosen centers 
in the multiple-LRG systems. Interestingly, 
the spectra for BLRG, FLRG and Mean centers all show
smaller amplitudes with increasing wavenumber, as expected in the FoG
effect. To be more precise, the power spectrum of FLRG center shows the
strongest FoG effect, because a larger fraction of FLRGs 
are satellite galaxies
than BLRGs (see Table \ref{tab:lrgs}). These results can be compared with Figure~2 in
\cite{Hikage12b}.
It can be found that the mock catalog qualitatively 
reproduces the SDSS measurements: the spectra of BLRG and Mean centers
are similar, and the spectrum for FLRG shows the stronger FoG
suppression. 

Figure~\ref{fig:pk} also shows the power spectrum using  the potential
minimum as the center of each halo.  We define the potential minimum as the
mass density peak of smooth component: the central
subhalo position, in each LRG-host halo. In this part of the analysis,
 the power spectrum is measured by using the position of a central
galaxy in each host halo. Again note that 
BLRG is not necessarily a central subhalo (galaxy) as shown in
Figure~\ref{blrg_sat}. The power spectrum for the potential minimum has
a smaller amplitude than that of the mass center of host halo, implying
that the potential minimum is moving around the mass center in
each halo.
Comparing the spectra
for the potential minimum and the BLRG center shows that the BLRG
spectrum has a smaller amplitude than the spectra for the potential
minimum or the mass center by a few \% in the fractional amplitude up to
$k\simeq 0.3~\Mpch$. The few \%-level FoG contamination would be
okay for a current accuracy of the 
power spectrum measurement, but will need to be
carefully taken into account for a higher-precision measurement of
upcoming redshift surveys. For comparison, the three dots-dashed curve shows
the effect on the real-space matter power spectrum caused by massive neutrinos, where we
assumed $m_{\nu,{\rm tot}}\simeq 0.1~$eV for
the total mass of neutrinos, close to the lower limit on the neutrino
mass for the inverted mass hierarchy. For the normal mass hierarchy, the
lower limit on the total mass is about 0.05~eV, and the amount of the
suppression is about half of the result of $0.1~$eV in
Figure~\ref{fig:pk}.
The lowest curve in the figure shows the difference of the real-space matter power spectrum when taking account of massive neutrinos relative to the spectrum for the mass-less neutrino cosmology.

In the right panel of Figure~\ref{fig:pk}, we also show the
redshift-space power spectrum derived by using  all the LRGs in the
catalog. Note that we properly subtracted the shot noise
contamination from the measured power spectra by using the number
densities of LRGs or LRG-host halos.  In this case, the power spectrum
ratio shows greater amplitudes with increasing wavenumber rather than
the FoG suppression. That is, the LRG power spectrum shows a greater
clustering power or greater bias than in the LRG-host halo spectrum.  
The scales shown here, the scales greater than a
few tens Mpc, are much larger than a virial radius of most massive host-halos
and the 1-halo term arising from clustering between two LRGs in the same host-halo
should not be significant at these scales.  
Hence, the greater amplitudes in the LRG
power spectrum would be due to 
 a more weight on more massive halos, because satellite LRGs
preferentially reside in more massive halos that have larger biases. 
Since the effect of
different linear bias should cause only a scale-independent change in the
power spectrum ratio, the change in the LRG power spectrum should be
from a stronger nonlinear bias of such massive halos, even though the FoG
suppression should be more significant
 for such halos. In fact, a combination of the
perturbation theory of structure formation and halo bias model seems to
reproduce such a non-trivial behavior in the power spectrum amplitudes
\citep{Nishizawaetal:12}. The results in the figure imply that
including satellite LRGs in the power spectrum analysis complicates
the interpretation of the measured power spectrum, thereby causing a
bias in cosmological parameters. These subtle effects need to be well
understood if we are going to use power spectrum measurements to place
unbiased constraints on cosmological parameters such as the
neutrino mass.

In Figure~\ref{fig:pk_halved}, we study how the residual FoG effect
varies with masses of LRG-host halos. To study this, we divide the LRG
halos into two sub-samples by masses of the LRG-halos smaller and larger
than the median, and measured the fractional power spectra for each
sub-sample relative to the halo sample. As expected, the FoG effect is
larger for the sub-sample containing more massive halos, because of the
higher fraction of satellite BLRGs as well as the larger velocity
dispersion (larger halo mass).  The right panel shows the similar
results, but obtained only by using the single-LRG halos. First of all,
the single LRG systems have a smaller FoG effect, because of the smaller
fraction of satellite BLRGs (Figure~\ref{blrg_sat} and Table \ref{tab:lrgs}) as well as the
smaller velocity dispersions for the lower-mass host-halos. 
Among the single-LRG halos, more massive
halos have relatively a larger FoG contamination, but only by a few
percent at $k\simlt 0.35~\Mpch$ in the amplitude. Thus, 
the
use of single-LRG systems may allow a cleaner interpretation of the
measured power spectrum, yielding a more robust, unbiased constraint on
cosmological parameters. 

\section{Discussion and Conclusion}
\label{sec:conclusion}

In this paper, we have developed a new abundance-matching based method to
generate a mock catalog of the SDSS LRGs, using catalogs of halos and
subhalos in $N$-body simulations. A brief summary of our method is as
follows: (1) identify LRG-progenitor halos at $z=2$ down to a certain
mass threshold until the comoving number density of the halos become
similar to that of the SDSS LRGs at $z=0.3$  (2) trace the
merging and assembly histories 
of the LRG ``star particles'', the  30\% innermost particles of
each 
$z=2$-LRG-progenitor halo that are gravitationally, tightly-bounded particles,
and (3) at $z=0.3$, identify the subhalos and halos hosting
the LRG "star" particles. If a subhalo at $z=0.3$ contains more than
50\% of 
the star particles of any progenitor halo, we assign 
the subhalo at $z=0.3$ as an LRG-host subhalo. 
We should emphasize that our method
does not employ {\em any} free fitting parameter to adjust 
in order for the
model to match the measurements, once
the mass threshold of the LRG-progenitor halos is determined to match the number density of SDSS LRGs.  Thus, by assuming that a majority of
stellar components of LRG is formed at $z=2$, we can trace the assembly
and merging histories of LRGs over a range of redshift, $z=[0.3,2]$; for
example, we can directly trace which LRGs become central or satellite
galaxies in the LRG-host halos at $z=0.3$. The novel aspect of our approach is that the
abundance matching of halos to a particular type of galaxies (LRGs in
this paper) is done by 
connecting the halos and subhalos at different
redshifts ($z=2$ and $z=0.3$ in our case), while the standard method is
done for the same or similar redshift to the redshift of target galaxies.
In addition, central and satellite subhalos are populated with galaxies under 
a single criterion: if a subhalo at $z=0.3$ is a descendant of the
$z2$-halos, the subhalo is included.
The standard abundance matching uses  different quantities for 
central and satellite subhalos, e.g., the circular velocities at the galaxy
redshift and at the accretion epoch, respectively. 

Using the mock catalog, we have computed various statistical quantities: the
halo occupation distribution, the projected correlation function of
LRGs, the mean surface mass density profile around LRGs (which is an
observable of the LRG-galaxy weak lensing), and the redshift-space power
spectrum of LRGs. We showed that the mock catalog predictions are in a
good agreement with the measurements from the SDSS LRG catalog
(Figures~\ref{hod}, \ref{wp}, \ref{fig:dsigma}, \ref{fig:dsigma_multi}
and \ref{fig:pk}). Thus our method seems to capture an essential feature
of LRG formation in  terms of a hierarchical structure formation
scenario of $\Lambda$CDM
model. 

In the SDSS sample,
about 5\%
of the halos contain multiple LRGs.  In our simulation, 8\%
of the halos contain multiple LRGs.  This modest deviation may be due
to our simplified assumptions. First, we assumed
that LRG progenitors are formed at a single epoch, $z=2$.
This is too simplified
assumption as LRG formation almost certainly took place over a range of redshifts. Including
a broader period of formation of LRG-progenitor halos may improve the model
prediction. Second, although LRGs are observationally selected by 
 magnitude and color cuts, our definition of
the LRG-progenitor halos at $z=2$ is solely based on their masses. The
agreements between our mock catalog and the SDSS measurements support
that the matching based on the LRG-progenitor halo masses seems fairly
reasonable to mimic a population of LRGs. However, the model may be
further refined by combining masses of the progenitor halos with other
indicators when matching to LRGs. For instance, using the maximum
circular velocity of each halo instead of its mass may improve the model
accuracy.
Another potential improvement would
be to introduce some stochasticity into the relationship between halo mass
and inclusion in the LRG sample.  Variation in star formation histories should
introduce scatter into the halo mass/galaxy luminosity relation.  We have
done a preliminary study where we  introduce some scatter and find 
that this lowers the characteristic halo mass, which leads to  smaller bias parameters,
and obviates the disagreement between theory and observation for the projected
correlation function or the lensing mass profile in Figures~\ref{wp} and
\ref{fig:dsigma}.
Another simplifying assumption was our neglect of  satellite subhalos in the
parent halo at $z=2$ in the abundance matching procedure. We naively expect
that subhalos at $z=2$ merge into central subhalos from $z=2$ to $z=0.3$
due to dynamical friction, so we used the simplest method as the first
attempt. However, including the subhalos at $z=2$ for the abundance
matching may improve an accuracy of the mock catalog. We plan to explore these
improvements in a future work.

Our mock catalog or more generally our abundance matching method offers
several applications to measurements. First, \cite{Masjedietal:08}
showed that, by using the small-scale clustering signal and the pair
counting statistics, LRGs are growing by about 1.7\% per Gyr, on
average from merger activity from $z=1$ to $z\sim 0.3$. Our method
directly traces how each LRG-progenitor halo acquires the
mass from other LRG-progenitor halos by major or minor mergers from
$z=2$ to $z=0.3$. Hence, we can compare the prediction of our mock
catalog with the measurement for the mass growth rate of LRGs. By using
the constraint, we may be able to further improve the mock catalog.

Second, our method can predict how the distribution of LRG-progenitor
evolves in relative to dark matter distribution as a function of
redshift. Thus, our mock catalog can be used to predict various
cross-correlations of LRG positions  with other tracers of
large-scale structure. As one such example, in this paper, we have studied
the LRG-galaxy weak lensing measured via cross-correlation of LRGs with
shapes of background galaxies, and have 
shown a remarkably good agreement
between our model and the SDSS measurements. Another cross-correlation
that has been studied in the literature is a cross-correlation of LRGs
with a map of cosmic microwave background (CMB) anisotropies, 
which probes the stacked
Sunyaev-Zel'dovich (SZ) effect \citep[][]{Handetal:11,Sehgaletal:12} or the lensing
effect on the CMB. Since every
massive halos always host at least one LRG 
 (100\% of halos with $M_{\rm vir}\ge2\times
10^{14}~\hMsun$ in our mock catalog), the cross-correlation is a
powerful cross-check of the SZ signals. Our mock catalog can predict how
the stacked SZ signals change for different catalogs of LRGs such as
an inclusion of satellite LRGs and multiple-LRG systems, which may be
able to resolve some tension between the observed LRG-CMB
cross-correlation signal and the theoretical expectation
\citep{Sehgaletal:12}. 

Third is an application of our method to LRGs or massive red galaxies
at higher redshift than $z=0.3$.  The SDSS-III BOSS survey is now
carrying out an even more massive redshift survey of SDSS imaging galaxies. 
The magnitude and color cuts used for the BOSS
survey are designed to efficiently select galaxies at $0.4<z<0.7$, and
are different from the SDSS-I/II LRG selection. The BOSS galaxies are
called the ``constant mass'' (CMASS) galaxies. The majority of CMASS
galaxies are early-type galaxies, but are not exactly the same
population as LRGs. In addition, the comoving number density of CMASS
galaxies is higher than that of LRGs by a factor of 3.
Hence, it would be interesting to apply the method developed in this
paper to the CMASS galaxies. Figure~\ref{fig:snapshot} shows an
interesting indication of our mock catalog: more LRG-progenitor halos
survive in the $z=0.5$ output than at $z=0.3$, because the halos do not
have enough time to experience merging due to the shorter time duration
from $z=2$ to $z=0.5$ than to $z=0.3$. 
Hence, our mock catalog naturally predicts a higher number
density of LRG-progenitor halos at higher redshift than at $z=0.3$, and
may be able to match some of the BOSS galaxies without any fine
tuning. 
Since the BOSS survey will provide us with
a higher-precision clustering measurement and therefore has the
potential to achieve 
tighter cosmological constraints, it is critically important to use an
accurate mock catalog of the CMASS galaxies in order to remove or
calibrate various systematic errors inherent in an unknown relation
between the CMASS galaxies and dark matter. We hope that our method is
useful for this purpose and can be used to attain the full potential of
the BOSS survey or more generally upcoming redshift surveys for
precision cosmology. This is our future study and will be presented
elsewhere.

\section*{Acknowledgments}
We thank Kevin Bundy, Surhud More and David Wake for useful discussion and valuable
comments. We also thank Rachel Mandelbaum, Zheng Zheng and Idit Zehavi
for kindly giving us their SDSS measurement results.
We appreciate Takahiro Nishimichi for kindly providing us with the second-order Lagrangian perturbation theory code.
We also thank  Naoki Yoshida for valuable advices for numerical 
techniques.
We are grateful to the anonymous referee for helpful comments.
SM acknowledges
support from a Japan Society for Promotion of Science (JSPS) fellowship 
(No. 23-6573).
This work was supported by JSPS Grant-in-Aid for Scientific Research
Numbers 22340056 (NS), 23340061 (MT), 24740160 (CH)
and NASA grant ATP11-0034 and ATP11-0090 (DS).
MT greatly thanks Department of Astrophysical Sciences, Princeton
University and members there for its warm hospitality during his visit,
where this work was done.  
The authors acknowledge Kobayashi-Maskawa Institute for the Origin of
Particles and the Universe, Nagoya University for providing computing
resources useful in conducting the research reported in this paper.
This research was in part supported by the Grant-in-Aid for Nagoya
University Global COE Program, "Quest for Fundamental Principles in the
Universe: from Particles to the Solar System and the Cosmos", from the Ministry
of Education, Culture, Sports, Science and Technology (MEXT) of Japan,
by
JSPS Core-to-Core Program ``International Research Network for Dark
Energy'', by Grant-in-Aid for Scientific Research from the JSPS
Promotion of Science, by Grant-in-Aid for Scientific Research on
Priority Areas No. 467 ``Probing the Dark Energy through an Extremely
Wide \& Deep Survey with Subaru Telescope'', by World Premier
International Research Center Initiative (WPI Initiative), MEXT, Japan,
by the FIRST program ``Subaru Measurements of Images and Redshifts
(SuMIRe)'', CSTP, Japan, and by the exchange program between JSPS and
DFG.

\bibliographystyle{mn2e} 
\label{lastpage}
\bibliography{mn-jour,refs}

\appendix
\section{Variants of our abundance matching method}
\label{app:parameters}

Although we have developed the simplest abundance matching method
by connecting halos at $z=2$ to subhalos at $z=0.3$, the method rests on
some simplified settings or assumptions that are not obvious: 
(1) the formation redshift of
LRG-progenitor halos is set to a single epoch of $z=2$, (2) 
we defined the ``LRG-star-particles'' by the innermost 30\% particles
of each $z=2$-progenitor halo, and (3) the ``matching'' fraction of the
star particles to each subhalo at $z=0.3$ is set to more than 50\% (if
a subhalo at $z=0.3$ contains more than 50\% of the star particles of
any $z=2$-progenitor halo, the subhalo is called as the LRG-host
subhalo). In summary, for our fiducial method, we have used 
%}
\begin{equation}
  z_{\rm form}=2, f_{\rm star}=0.3,~~{\rm and}~~f_{\rm match}=0.5,
\end{equation}
as described in detail in Section~\ref{sec:method}. In this Appendix, we
study how the results are changed if varying each of these parameters to
other values. In doing this, we use the L300 run because it has a higher
resolution than L1000 run and allows us 
to better resolve less massive
halos or sub-halos from the $z=2$ or $z=0.3$ simulation outputs.

\subsection{LRG-progenitor halo formation redshift: $z_{\rm form}$}

\begin{figure}
  \includegraphics[width=8.5cm]{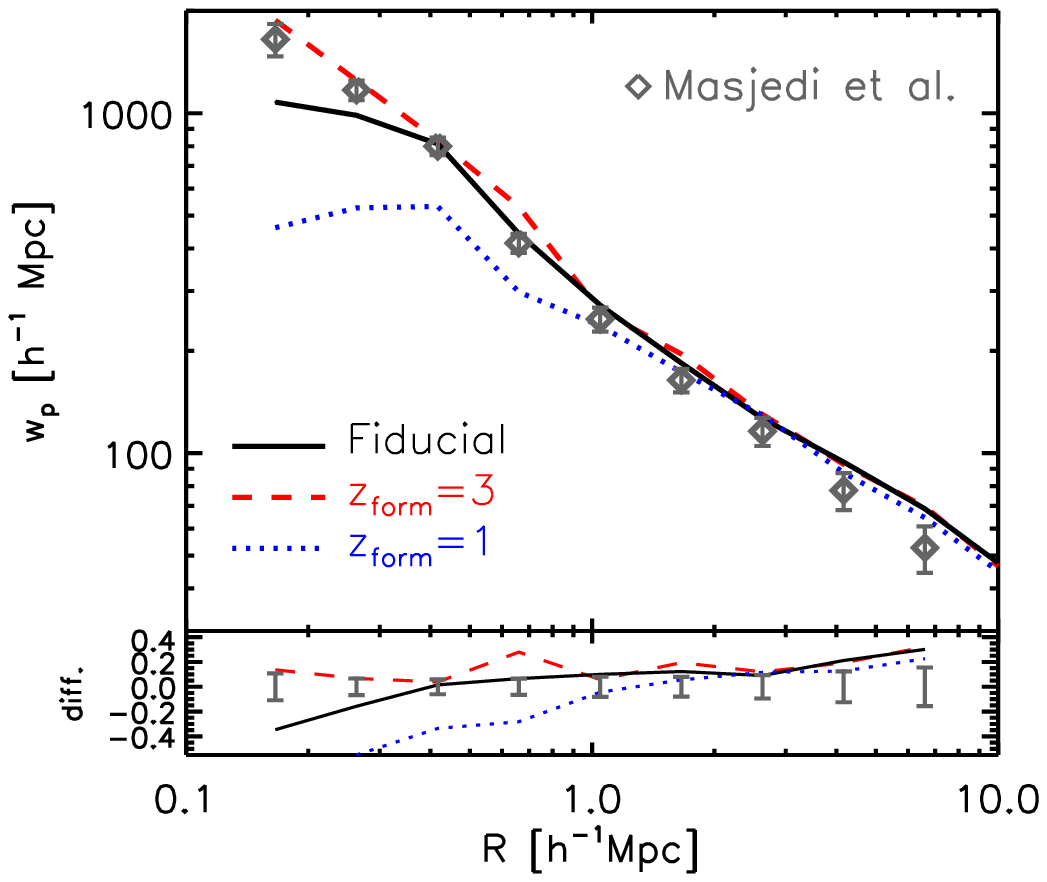}
  \caption{Shown is how the model prediction of the projected
 correlation function is changed if varying the model
 parameter in our abundance matching method, in comparison with the
 SDSS measurement as in Figure~\ref{wp}. 
This plot shows the correlation functions from the mock
 catalogs obtained by changing the formation redshift of LRG-progenitor
 halos from $z_{\rm form}=2$ (our fiducial choice) to $z_{\rm form}=1$
 or $3$. The different formation redshifts change the prediction mainly at
 the small scales, in the one-halo regime, because it changes a time duration
 for each progenitor halo to experience subsequent merging and
 assembly histories. For example, if changing to $z_{\rm form }=1$, the
 progenitor halos do not have enough time to experience merger, which
 decreases a population of satellite LRGs (subhalos) and in turn leads
 to the decreased clustering power at the small scales.
}
  \label{wp_zform}
\end{figure}
\begin{figure}
  \includegraphics[width=8.5cm]{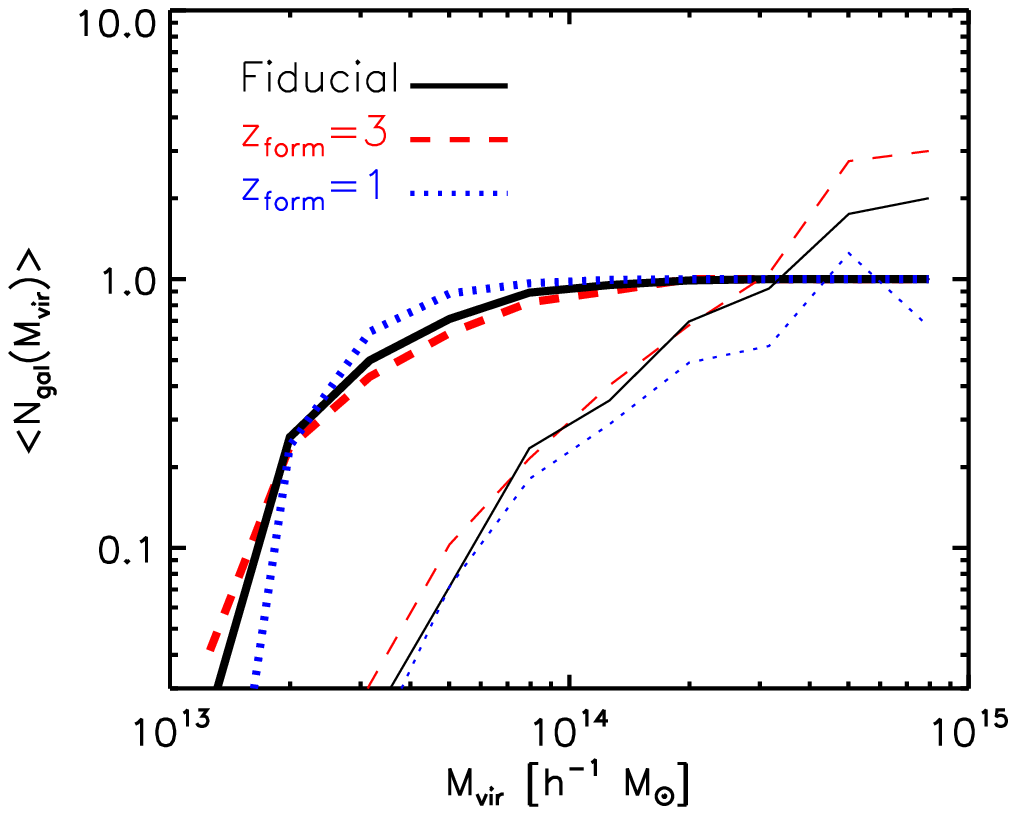}
  \includegraphics[width=8.5cm]{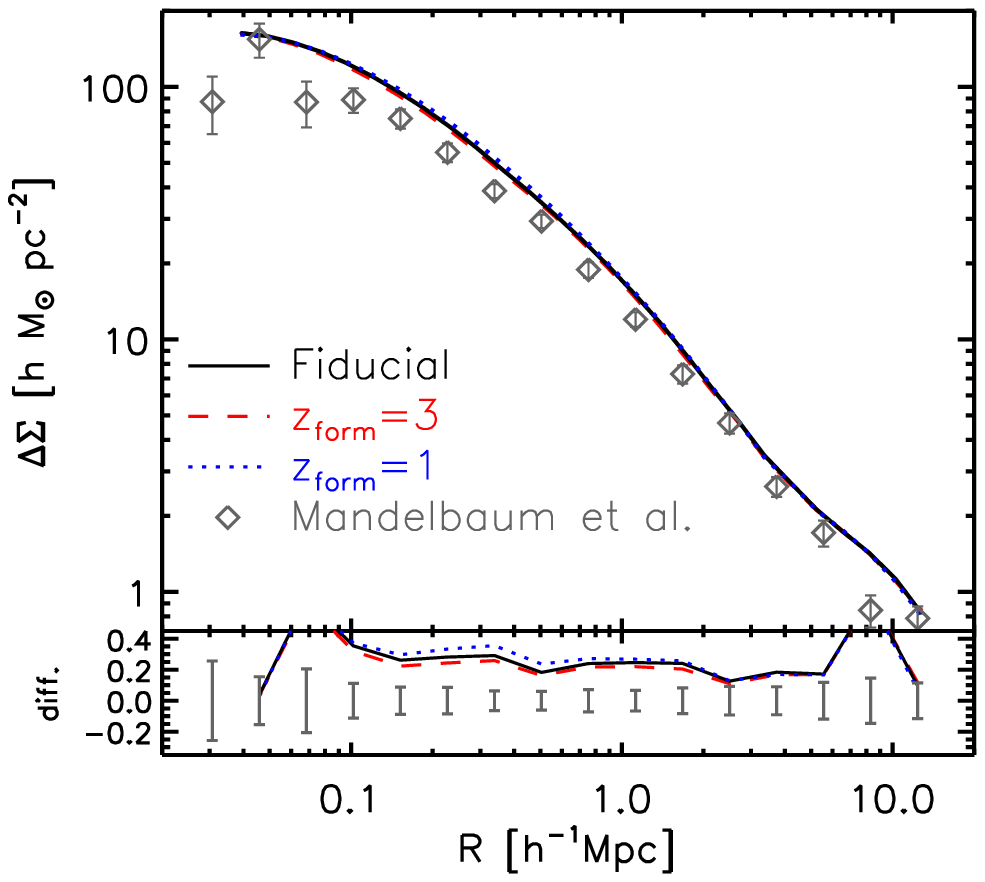}
\caption{Similarly to the previous plot, but 
for the HOD ({\em upper panel}) and for the projected mass profile 
({\em lower}).
\label{fig:hod_zform}} 
\end{figure}

Throughout this paper, we employed $z_{\rm form}=2$ for the
formation redshift of LRG-progenitor halos, as the first attempt,
motivated by the fact that LRGs typically have old ages $\simgt 5~{\rm
Gyr}$.  Figure~\ref{wp_zform} shows how the mock catalog of LRGs is
changed if varying the formation redshift to $z_{\rm form}=1$ or $3$ from
our fiducial choice $z_{\rm form}=2$. Here, to assess the difference, we
show the projected correlation functions obtained from the mock
catalogs. The figure shows that the formation redshift of $z_{\rm
form}=1$ or $3$ gives lower or higher amplitudes at small separations
$\simlt 1~\hMpc$, which is
in the one-halo term
regime, than our fiducial case of $z_{\rm form}=2$.  Nevertheless,
the encouraging result is that the large-separation correlation function
in the two-halo regime is not largely changed.  These changes can be
understood as follows. For the case of $z_{\rm form}=1$, the
LRG-progenitor halos have a shorter duration since the
formation and each halo has a smaller chance to experience
subsequent mergers than in the case of $z_{\rm form}=2$, the progenitor
in turn has a smaller chance to be included in multiple-LRG systems
being a satellite subhalo at $z=0.3$, which decreases 
clustering power in the one-halo regime.  Thus Figure~\ref{wp_zform}
implies that a choice of $z_{\rm form}\simgt 2$ is reasonable.

The upper panel of Figure~\ref{fig:hod_zform} shows how the change of
$z_{\rm form}$ alters the HOD as in Figures~\ref{hod}. As we discussed
above, the change to $z_{\rm form}=3$ from $z_{\rm form}=2$ leads to a
smoother HOD shape around the cutoff halo mass for the central LRG HOD
extending down to less massive halos as well as to an increase of
satellite LRGs. The choice of $z_{\rm form}=1$ leads to opposite
effects.  On the other hand, the lower panel shows that the projected
mass profile of LRG-host halos is almost unchanged by the change of
$z_{\rm form}$.  Thus an accuracy of the mock catalog can be improved,
especially for predicting the small-scale clustering, by further
introducing a variation of the formation epochs as additional parameter.

\subsection{The fraction of LRG-star particles in $z=2$-LRG-progenitor
  halo: $f_{\rm star}$}

\begin{figure}
  \includegraphics[width=8.5cm]{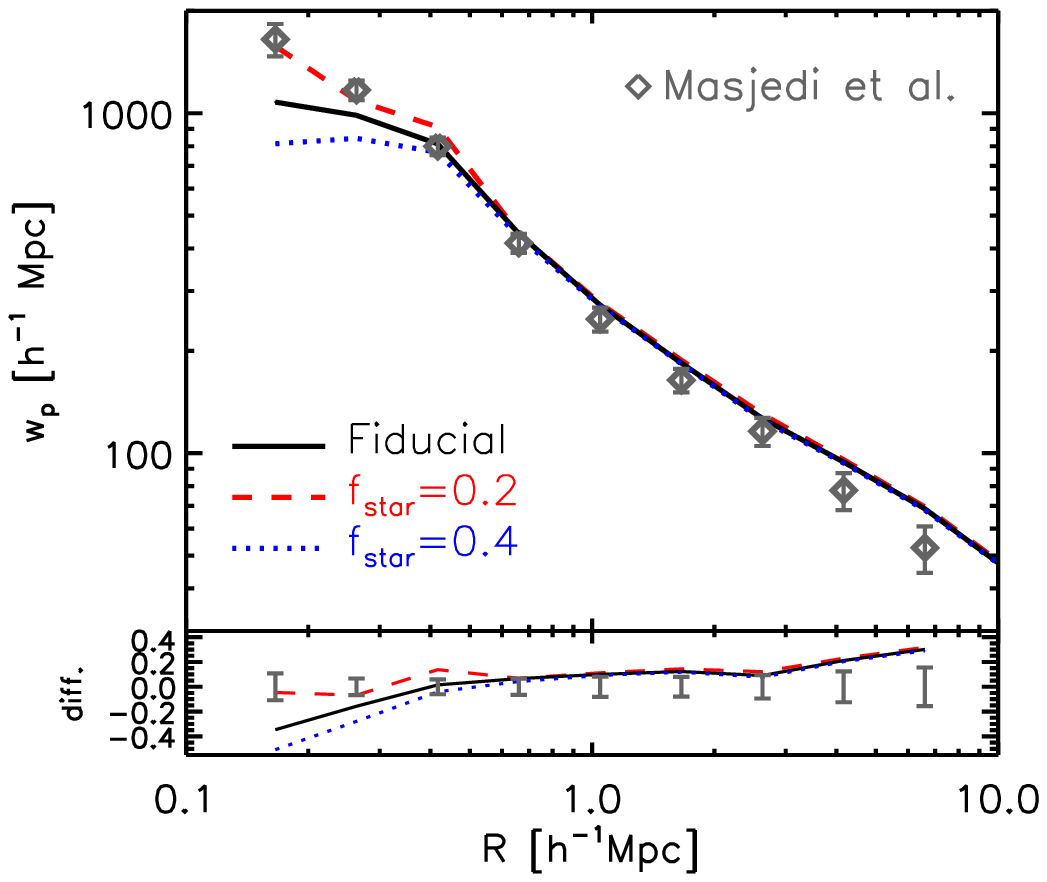}
  \includegraphics[width=8.5cm]{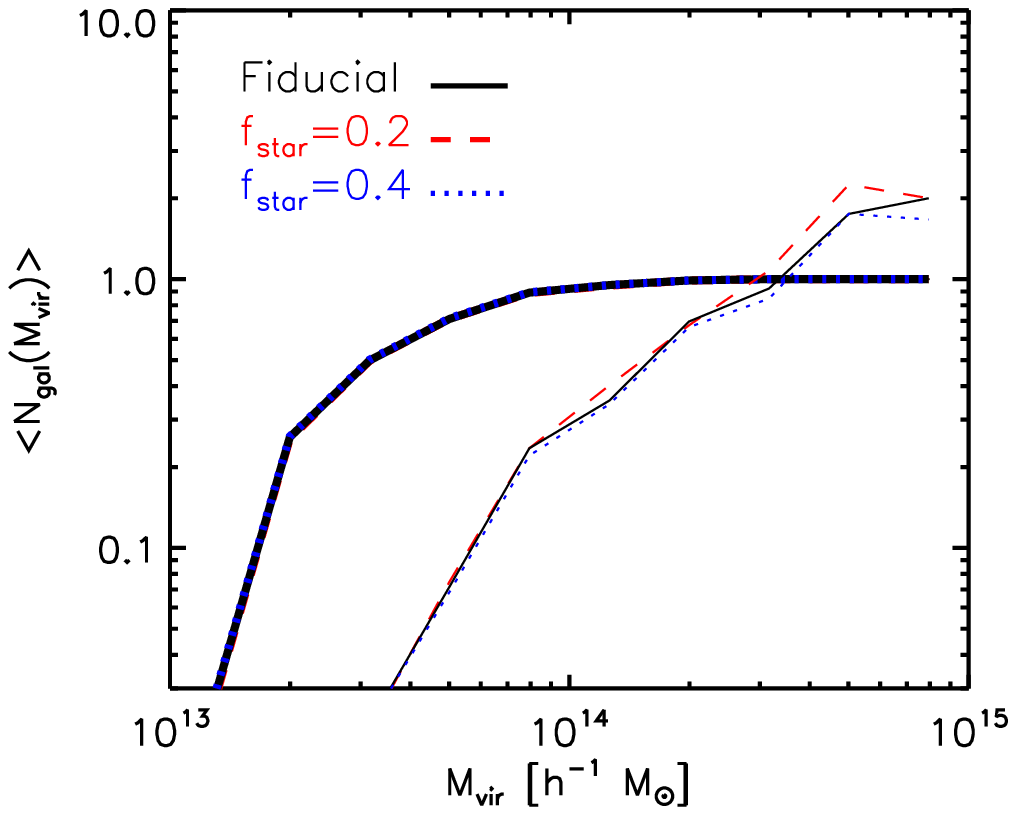}
  \caption{
Similarly to the previous figure, but this plot shows the
 impact of the parameter $f_{\rm star}$ on the correlation function
 prediction ({\em upper panel}) and on the HOD ({\em lower}),
where $f_{\rm star}$ is used to define ``LRG-star''
 particles in each LRG-progenitor halo at $z=2$ by the 
 $f_{\rm star}$-fraction of innermost member particles in the halo. 
Again, changing $f_{\rm star}=0.2$ or $0.4$ from our fiducial
 choice $f_{\rm star}=0.3$ alters the model prediction at the small
 scales. 
}
  \label{wp_BndFrac}
\end{figure}

To trace the LRG-progenitor halos from $z=2$ to $z=0.3$, we defined the
LRG-star particles in each $z=2$
LRG-progenitor halo
by the $30\%$ innermost particles of the FoF member particles, 
$f_{\rm star}=0.3$,
 assuming that the stars are formed
at the central region of each halo. However, the fraction 30\%
is an
arbitrary choice. 

Figure~\ref{wp_BndFrac} shows how the projected correlation and the HOD
are changed by varying to $f_{\rm star}=0.2$ or $0.4$ from the fiducial
choice of $f_{\rm star}=0.3$. Note that, when matching each
$z=2$-progenitor halo to a subhalo at $z=0.3$, we imposed the condition
that a subhalo at $z=0.3$ should have more than 50\% of the star
particles in each of the mock catalogs.  Figure~\ref{wp_BndFrac} shows
that the change of $f_{\rm star}$ alters the correlation function at the
small separation, in the one-halo regime. This is also found from the
lower panel, which shows that the change alters the satellite HOD.  If
we use $f_{\rm star}=0.4$ from $0.3$, it tends to include, in the star
particles, more loosely-bounded member particles in each
$z=2$-progenitor halo, the star particles tend to be stripped by tidal
interaction or merger with other halos, and then it is in turn difficult
to satisfy the 50\% matching condition to $z=0.3$ subhalo. Thus this
results in a smaller number of the survived satellite LRG-halos at
$z=0.3$, which causes to reduce the correlation amplitude at the small
scales. The opposite is true for the case of $f_{\rm star}=0.2$, because
it leads to a larger number of the survived satellite LRG-halos at
$z=0.3$ than in the case of $f_{\rm star}=0.3$. Nevertheless, the
encouraging result is the correlation function at large scales in the
two-halo regime is not sensitive to the variation of $f_{\rm star}$.
We have found that the projected mass profile is not changed as in 
the lower panel of Figure~\ref{fig:hod_zform}.

\subsection{The matching fraction between $z=2$-LRG-progenitor halos and 
 $z=0.3$-subhalos: $f_{\rm match}$}

\begin{figure}
  \includegraphics[width=8.5cm]{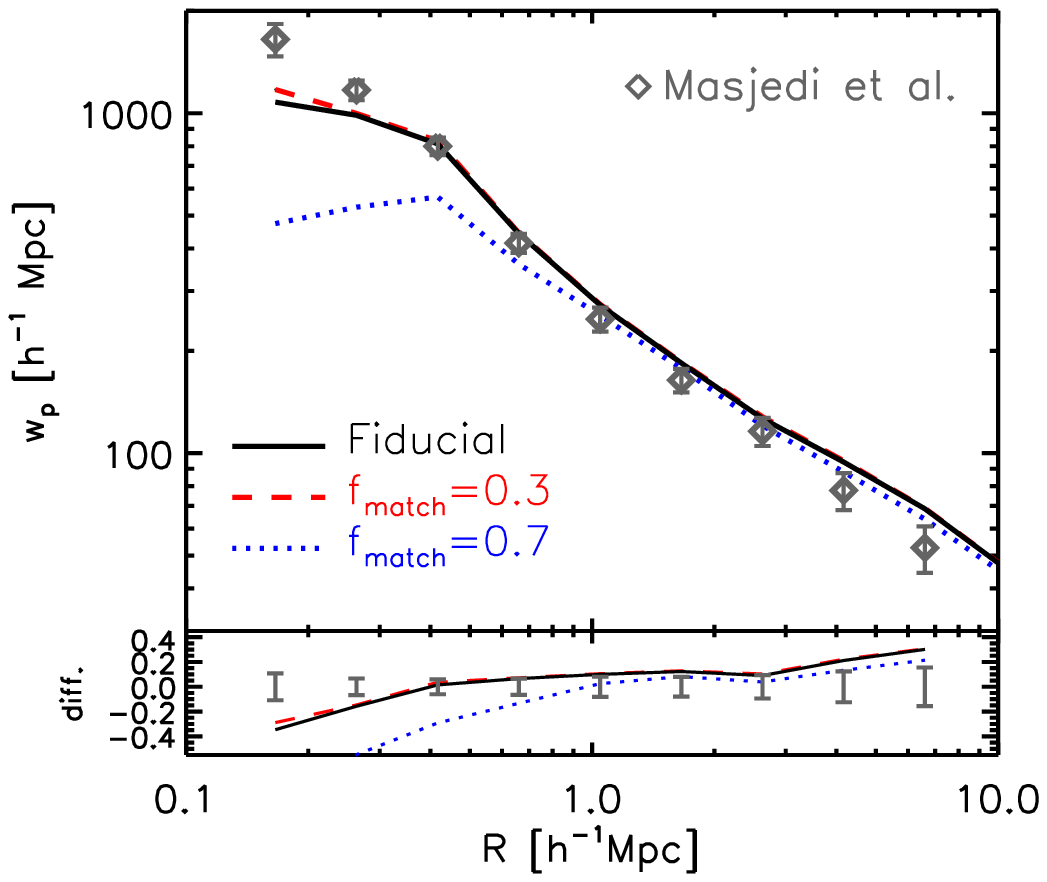}
  \includegraphics[width=8.5cm]{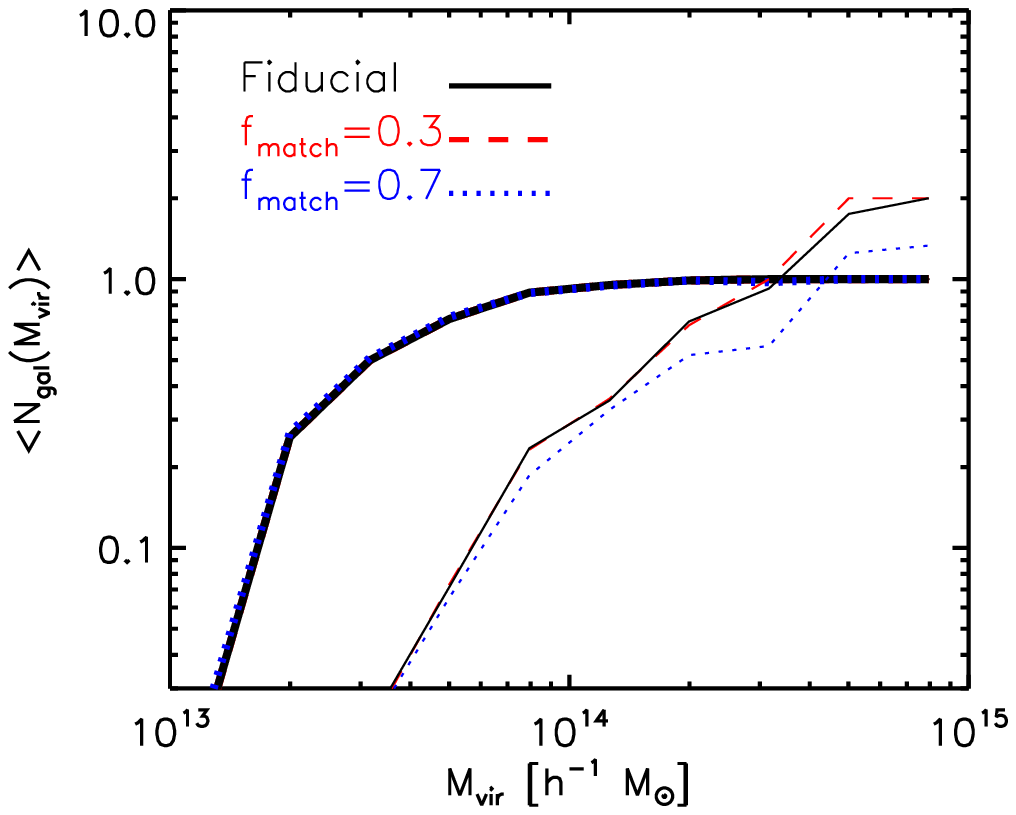}
  \caption{Similarly to the previous figure, but this plot show the
    impact of the matching fraction parameter $f_{\rm match}$. In the case of
    $f_{\rm match }=0.5$ (our fiducial choice), a subhalo at $z=0.3$ is called as
    LRG-host subhalo if the subhalo contains more than 50\% of the
    star particles in an LRG-progenitor halo at $z=2$. The different curves
    show the results for $f_{\rm match}=0.3$ or $0.7$, which differ from our
    fiducial model at the small scales.  
  }
 \label{wp_rmatch}
\end{figure}

Finally, we study how the mock catalog is changed by varying the
matching fraction to $f_{\rm match}=0.3$ or $0.7$ from our fiducial choice 
$f_{\rm match}=0.5$.
Figure~\ref{wp_rmatch} shows that the change of $f_{\rm
match}$ alters the projected correlation function at small scales in the
one-halo regime and the satellite HOD, 
similarly to Figures~\ref{wp_zform} 
and \ref{wp_BndFrac}. Again, if increasing the matching fraction to
$f_{\rm match}=0.7$ from $0.5$, it leads to a less number of the matched
satellite subhalos at $z=0.3$ and in turn leads to a decreased power in
the correlation function. On the other hand, the large-scale correlation
is more robust to the change of $f_{\rm match}$ similarly to the
previous figures.
We have again found that the projected mass profile is not changed as in 
the lower panel of Figure~\ref{fig:hod_zform}.

\end{document}